\documentclass[aps,pra,reprint,groupedaddress]{revtex4-1}
\usepackage{amssymb}
\usepackage{amsmath}
\usepackage{array}
\usepackage{multirow,bigdelim}
\usepackage{enumitem}
\usepackage{graphicx}
\usepackage{epsfig}
\usepackage{graphics}
\usepackage{latexsym}

\begin{document}

\title{ Entanglement classification via integer partitions }
\author{ Dafa Li$^{1,2}$}

\begin{abstract}
Abstract:

In [M. Walter et al., Science 340, 1205, 7 June (2013)], they gave a
sufficient condition for genuinely entangled pure states and discussed SLOCC
classification via polytopes and the eigenvalues of the single-particle
states. In this paper, for $4n$ qubits, we show the invariance of algebraic
multiplicities (AMs) and geometric multiplicities (GMs) of eigenvalues and
the invariance of sizes of Jordan blocks (JBs) of the coefficient matrices
under SLOCC. We explore properties of spectra, eigenvectors, generalized
eigenvectors, standard Jordan normal forms (SJNFs), and Jordan chains of the
coefficient matrices. The properties and invariance permit a reduction of
SLOCC classification of $4n$ qubits to integer partitions (in number theory)
of the number $2^{2n}-k$ and the AMs.
\end{abstract}

\affiliation{
$^1$Department of Mathematical Sciences, Tsinghua University,
Beijing, 100084, China\\
$^2$Center for Quantum Information Science and Technology, Tsinghua National
Laboratory for Information Science and Technology (TNList), Beijing,
100084, China\\
email address: lidafa@tsinghua.edu.cn}

\maketitle

\section{ Introduction}

As the subtle properties of entangled states are applied in quantum
information and computation, many efforts have contributed to understanding
the different ways of entanglement \cite{Nielsen}. Clearly, local quantum
operations cannot change the non-local properties of a state. The
entanglement for two and three qubits are well known. However, it is hard to
classify multipartite entanglement for four or more qubits. To reach the
purpose, SLOCC equivalence of two states of a multipartite system was
proposed and formulated \cite{Bennett, Dur}. It is known that two states in
the same SLOCC equivalence class can do the same tasks of quantum
information theory, although with different success probabilities \cite{Dur,
Verstraete, Horodecki}.

D\"{u}r \emph{et al.} classified three qubits into six SLOCC classes
including the classes GHZ and W, and indicated that there are an infinite
number of SLOCC\ classes for four or more qubits. In the pioneering work
\cite{Verstraete}, Verstraete \emph{et al.} classified the infinite number
of SLOCC\ classes of four qubits into nine families under determinant one
SLOCC by using a generalization of\ the singular value decomposition. After
then, SLOCC\ classification of four qubits were studied deeply \cite{Miyake,
Luque, LDF07b,Chterental,Lamata,LDFQIC09,Borsten, Viehmann, Ribeiro, Buniy,
Sharma12}.

For SLOCC classification of n qubits, the previous articles proposed
different SLOCC\ invariants: for example, the concurrence and 3-tangle \cite%
{Coffman}; local ranks for three qubits \cite{Dur}; polynomial\ invariants
\cite{Wong, Miyake, Luque, Leifer, Levay, LDF07a, Djokovic, LDFQIP,
Osterloh05, Viehmann, LDFJPA13, Gour, LDFPRA13} of which the invariant
polynomials of degrees 2 for even n qubits \cite{LDF07a}, 4 for $n\geq 4$
(odd and even) qubits \cite{LDF07a, LDFPRA13}, and 6 for even $n\geq 4$
qubits \cite{LDFJPA13}; the diversity degree and the degeneracy
configuration of a symmetric state \cite{Bastin}; ranks of coefficient
matrices \cite{LDFPRL12,LDFPRA12, FanJPA, Wang, LDFPRA15}; the entanglement
polytopes \cite{Walter}. Recently, spectra and SJNFs of 4 by 4 matrices were
used to investigate SLOCC\ classification of\ pure states of n qubits \cite%
{LDFQIP18}.

In this paper, we show the invariance of algebraic and geometric
multiplicities of eigenvalues and sizes of JBs under SLOCC\ for $4n$ qubits.
We investigate properties of spectra, eigenvectors, generalized
eigenvectors, SJNFs, and Jordan chains of matrices $\Phi _{2^{2n+1}}$. Via
integer partitions, the properties, and the invariance, we classify pure
states of $4n$ qubits, specially four qubits, under SLOCC.

This paper is organized as follows. In Section 2, we show the invariance of
algebraic and geometric multiplicities of eigenvalues and sizes of JBs under
SLOCC\ for $4n$ qubits. In Section 3, via integer partitions,\ we classify
spectra of $\Phi _{2^{2n+1}}$ and pure states of $4n$ qubits. In Section 4,
via integer partitions,\ we classify SJNFs of $\Phi _{2^{2n+1}}$ and pure
states of $4n$ qubits.

\section{\textbf{Invariant AMs, GMs, and sizes of JBs under SLOCC}}

Let $|\psi \rangle =\sum_{i=0}^{2^{4n}-1}a_{i}|i\rangle $ be any pure state
of $4n$ qubits, where $a_{i}$ are coefficients. It is well known that two $%
4n $-qubit pure states $|\psi \rangle $ and $|\psi ^{\prime }\rangle $ are
SLOCC equivalent if and if there is an invertible local operator $\mathcal{A}%
_{1}\otimes \cdots \otimes \mathcal{A}_{4n}$ such that
\begin{equation}
|\psi ^{\prime }\rangle =\mathcal{A}_{1}\otimes \mathcal{A}_{2}\otimes
\cdots \otimes \mathcal{A}_{4n}|\psi \rangle ,
\end{equation}%
where $\mathcal{A}_{i}\in CL(C,2)$ \cite{Dur}.

Let $C_{q_{1}q_{2}\cdots q_{2n}}(|\psi \rangle )$ be the coefficient matrix
of the state $|\psi \rangle $ of $4n$ qubits, i.e. entries of the matrix are
the coefficients of the state $|\psi \rangle $, where $q_{1}$, $q_{2}$, $%
\cdots $, and $q_{2n}$ are chosen as row bits while $q_{2n+1}$, $q_{2n+2}$, $%
\cdots $, and $q_{4n}$ are chosen as column bits. Clearly, $%
C_{q_{1}q_{2}\cdots q_{2n}}$ is a $2^{2n}$ by $2^{2n}$ matrix.

It is known that for any two SLOCC equivalent pure states $|\psi \rangle $
and $|\psi ^{\prime }\rangle $ of $4n$ qubits, the matrices $%
C_{q_{1}q_{2}\cdots q_{2n}}$ satisfy the following equation \cite{LDFPRL12}%
\cite{LDFPRA12}\cite{LDFPRA15},
\begin{equation}
C_{q_{1}q_{2}\cdots q_{2n}}(|\psi ^{\prime }\rangle )=\Delta
_{1}C_{q_{1}q_{2}\cdots q_{2n}}(|\psi \rangle )\Delta _{2},  \label{reduce-2}
\end{equation}%
where $\Delta _{1}=(\mathcal{A}_{q_{1}}\otimes \mathcal{A}_{q_{2}}\otimes
\cdots \otimes A_{q_{2n}})$ and $\Delta _{2}=(\mathcal{A}_{q_{2n+1}}\otimes
\cdots \otimes \mathcal{A}_{q_{4n}})^{t}$. Note that $\mathcal{A}^{t}$ is
the transpose of $\mathcal{A}$.

Let%
\begin{equation}
T=\frac{1}{\sqrt{2}}\left(
\begin{array}{cccc}
1 & 0 & 0 & 1 \\
0 & i & i & 0 \\
0 & -1 & 1 & 0 \\
i & 0 & 0 & -i%
\end{array}%
\right)
\end{equation}%
and
\begin{equation}
U=T^{\otimes n}.  \label{u-mat-1}
\end{equation}%
It is easy to see that $T$ and $U$ are unitary. We make a conjugation of $%
C_{q_{1}q_{2}\cdots q_{2n}}(|\psi \rangle )$ by the unitary matrix $U$ in
Eq. (\ref{u-mat-1}) as follows. Let
\begin{equation}
\Gamma _{2^{2n}}(|\psi ^{\prime }\rangle )=UC_{q_{1}q_{2}\cdots
q_{2n}}(|\psi ^{\prime }\rangle )U^{+}
\end{equation}%
and
\begin{equation}
\Gamma _{2^{2n}}(|\psi \rangle )=UC_{q_{1}q_{2}\cdots q_{2n}}(|\psi \rangle
)U^{+}.
\end{equation}%
Let $Q_{1}=U\Delta _{1}U^{+}$ and $Q_{2}=U\Delta _{2}U^{+}$. From Eq. (\ref%
{reduce-2}), we obtain

\begin{eqnarray}
&&UC_{q_{1}q_{2}\cdots q_{2n}}(|\psi ^{\prime }\rangle )U^{+}  \notag \\
&=&U\Delta _{1}C_{q_{1}q_{2}\cdots q_{2n}}(|\psi \rangle )\Delta _{2}U^{+} \\
&=&U\Delta _{1}U^{+}UC_{q_{1}q_{2}\cdots q_{2n}}(|\psi \rangle )U^{+}U\Delta
_{2}U^{+}
\end{eqnarray}%
and then

\begin{equation}
\Gamma _{2^{2n}}(|\psi ^{\prime }\rangle )=Q_{1}\Gamma _{2^{2n}}(|\psi
\rangle )Q_{2}.  \label{simi-1}
\end{equation}%
Clearly, $\Gamma _{2^{2n}}(|\psi ^{\prime }\rangle )$ is not similar to $%
\Gamma _{2^{2n}}(|\psi \rangle )$.

Let us consider the matrix
\begin{equation}
\Phi _{2^{2n+1}}(|\psi \rangle )=\left(
\begin{array}{cc}
& \Gamma _{2^{2n}}(|\psi \rangle ) \\
\lbrack \Gamma _{2^{2n}}(|\psi \rangle )]^{t} &
\end{array}%
\right) .  \label{coef-1}
\end{equation}%
Via Eq. (\ref{simi-1}), a calculation derives
\begin{equation}
\Phi _{2^{2n+1}}(|\psi ^{\prime }\rangle )=O\Phi _{2^{2n+1}}(|\psi \rangle
)O^{t},  \label{P-S-1}
\end{equation}%
where
\begin{equation}
O=\left(
\begin{array}{cc}
Q_{1} &  \\
& Q_{2}^{t}%
\end{array}%
\right) .
\end{equation}%
Clearly,
\begin{equation}
O^{t}O=\left(
\begin{array}{cc}
Q_{1}^{t}Q_{1} &  \\
& Q_{2}Q_{2}^{t}%
\end{array}%
\right) =\left(
\begin{array}{cc}
gI_{2^{2n}} &  \\
& hI_{2^{2n}}%
\end{array}%
\right) ,  \label{P-S-2}
\end{equation}%
where
\begin{equation}
g=\Pi _{i=1}^{2n}\det A_{q_{i}}
\end{equation}%
and

\begin{equation}
h=\Pi _{i=1}^{2n}\det A_{q_{2n+i}}
\end{equation}
from Eqs. (\ref{orth-1}, \ref{orth-2}) in Appendix A.

Note that neither $Q_{i}$ nor $O$\ is orthogonal except that $\mathcal{A}%
_{i}\in SL(C,2)$. Therefore, SLOCC cannot guarantee that $\Phi
_{2^{2n+1}}(|\psi ^{\prime }\rangle )$ and $\Phi _{2^{2n+1}}(|\psi \rangle )$
are similar. Anyway, from Eqs. (\ref{P-S-1}, \ref{P-S-2}) we obtain%
\begin{eqnarray}
\Phi _{2^{2n+1}}(|\psi ^{\prime }\rangle ) &=&O\Phi _{2^{2n+1}}(|\psi
\rangle )O^{t}OO^{-1}  \notag \\
&=&O\Theta O^{-1},  \label{P-S-3}
\end{eqnarray}%
where
\begin{eqnarray}
\Theta &=&\Phi _{2^{2n+1}}(|\psi \rangle )O^{t}O  \notag \\
&=&\Phi _{2^{2n+1}}(|\psi \rangle )\left(
\begin{array}{cc}
gI_{2^{2n}} &  \\
& hI_{2^{2n}}%
\end{array}%
\right)  \notag \\
&=&\left(
\begin{array}{cc}
& h\Gamma _{2^{2n}}(|\psi \rangle ) \\
g[\Gamma _{2^{2n}}(|\psi \rangle )]^{t} &
\end{array}%
\right) .  \label{P-S-4}
\end{eqnarray}

In general, a square complex matrix $M$ is similar to a block diagonal
matrix
\begin{equation}
J=\left(
\begin{array}{cccc}
J_{i_{1}}(\lambda _{1}) &  &  &  \\
& J_{i_{2}}(\lambda _{2}) &  &  \\
&  & \ddots &  \\
&  &  & J_{i_{m}}(\lambda _{m})%
\end{array}%
\right) ,
\end{equation}%
where

\begin{equation}
J_{i_{l}}(\lambda _{l})=\left(
\begin{array}{cccc}
\lambda _{l} & 1 &  &  \\
& \lambda _{l} & \ddots &  \\
&  & \ddots & 1 \\
&  &  & \lambda _{l}%
\end{array}%
\right)
\end{equation}%
is a standard Jordan block with the eigenvalue $\lambda _{l}$, where $i_{l}$%
\ is the size of the block. Usually, $J$ is written as the direct sum
\begin{equation}
J=J_{i_{1}}(\lambda _{1})\oplus J_{i_{2}}(\lambda _{2})\oplus \cdots \oplus
J_{i_{m}}(\lambda _{m})
\end{equation}
of the Jordan blocks $J_{i_{1}}(\lambda _{1})$, $J_{i_{2}}(\lambda _{2})$, $%
\cdots $, and $J_{i_{m}}(\lambda _{m})$. In this paper, we write the direct
sum as

\begin{equation}
J=J_{i_{1}}(\lambda _{1})J_{i_{2}}(\lambda _{2})\cdots J_{i_{m}}(\lambda
_{m})  \label{direct-1}
\end{equation}%
by omitting \textquotedblleft $\oplus $\textquotedblright . We call Eq. (\ref%
{direct-1}) the SJNF\ of the matrix $M$.

In this paper, we define that two SJNFs
\begin{equation}
J_{i_{1}}(\beta _{1})J_{i_{2}}(\beta _{2})\cdots J_{i_{k}}(\beta _{k})
\end{equation}%
and
\begin{equation}
J_{i_{1}}(\eta \beta _{1})J_{i_{2}}(\eta \beta _{2})\cdots J_{i_{k}}(\eta
\beta _{k}),
\end{equation}
where $\eta \neq 0$, are proportional. For example, the SJNFs $%
J_{1}(1)J_{2}(1)J_{3}(2)$ and $J_{1}(3)J_{2}(3)J_{3}(6)$ are proportional.

Though we cannot guarantee that $\Phi _{2^{2n+1}}(|\psi ^{\prime }\rangle )$
and $\Phi _{2^{2n+1}}(|\psi \rangle )$ are similar, we can next show that
their spectra and SJNFs are proportional.

From Eq. (\ref{P-S-4}), we have the following result.

\textit{Lemma 1. }Spectra and SJNFs of $\Theta $ and $\Phi _{2^{2n+1}}(|\psi
\rangle )$ are proportional.

Property 1 in Appendix B means that their spectra are proportional.

We next show that SJNFs of $\Theta $ and $\Phi _{2^{2n+1}}(|\psi \rangle )$
are proportional. From the linear algebra, for any JB $J_{r}(\lambda )$ of $%
\Phi _{2^{2n+1}}(|\psi \rangle )$, $\Phi _{2^{2n+1}}(|\psi \rangle )$ has a
Jordan chain $v_{i}$, $i=1,2,\cdots ,r$, where $v_{1}$ is the eigenvector of
$\Phi _{2^{2n+1}}(|\psi \rangle )$ corresponding to the eigenvalue $\lambda $%
. Here, let $v_{i}$ be column vectors $\left(
\begin{array}{c}
v_{i}^{\prime } \\
v_{i}^{\prime \prime }%
\end{array}%
\right) $, where two blocks $v_{i}^{\prime }$ and $v_{i}^{\prime \prime }$
are of the same size. In light of Property 3 in Appendix B, we can construct
a chain of $\Theta $: $z_{i}$, $i=1,2,\cdots ,r$, where $z_{1}=\left(
\begin{array}{c}
\sqrt{h/g}v_{1}^{\prime } \\
v_{1}^{\prime \prime }%
\end{array}%
\right) $, which is the eigenvector of $\Theta $ corresponding to the
eigenvalue $\sqrt{gh}\lambda $. For the chain $z_{i}$, from the linear
algebra $\Theta $ has the JB $J_{r}^{\prime }(\sqrt{gh}\lambda )$.

Let the generalized modal matrices $M$ and $M^{\prime }$\ of $\Phi
_{2^{2n+1}}(|\psi \rangle )$ and $\Theta $ consist entirely of Jordan chains
of $\Phi _{2^{2n+1}}(|\psi \rangle )$ and $\Theta $, respectively, and let
the Jordan chains $v_{i}$ and $z_{i}$ be the $kth$ Jordan chains of $M$ and $%
M^{\prime }$, respectively. Then, the JBs $J_{r}(\lambda )$ and $%
J_{r}^{\prime }(\sqrt{gh}\lambda )$ are the $kth$ JBs of $\Phi
_{2^{2n+1}}(|\psi \rangle )$ and $\Theta $, respectively. Therefore, SJNFs
of $\Theta $ and $\Phi _{2^{2n+1}}(|\psi \rangle )$ are proportional.

Thus,\ Eq. (\ref{P-S-3}) and Lemma 1 lead to the following theorem.

\textit{Theorem 1.} If the states $|\psi \rangle $ and $|\psi ^{\prime
}\rangle $ of $4n$\ qubits\ are SLOCC equivalent, then spectra and SJNFs of $%
\Phi _{2^{2n+1}}(|\psi ^{\prime }\rangle )$ and $\Phi _{2^{2n+1}}(|\psi
\rangle )$ are proportional, respectively.

The following is our argument. Eq. (\ref{P-S-3}) implies that $\Phi
_{2^{2n+1}}(|\psi ^{\prime }\rangle )$ and $\Theta $ are similar. Therefore,
$\Phi _{2^{2n+1}}(|\psi ^{\prime }\rangle )$ and $\Theta $ have the same
spectra and SJNFs (ignoring the order of JBs). In light of Lemma 1, $\Theta $
and $\Phi _{2^{2n+1}}(|\psi \rangle )$ have the proportional spectra and
SJNFs.

Restated in the contrapositive the theorem reads: If spectra or SJNFs of\
two matrices $\Phi _{2^{2n+1}}(|\psi \rangle )$\ in Eq. (\ref{coef-1})
associated with two $4n$-qubit pure states are not proportional, then the
two states are SLOCC inequivalent.

For example, for four qubits, let $|\Upsilon \rangle =\sum_{i,j,k,l\in
\{0,1\}}|ijkl\rangle -|0000\rangle -|1111\rangle $. In light of Theorem 1,
one can test that $|\Upsilon \rangle $ is inequivalent to the states GHZ, W,
Cluster, or the Dicke state $|2,4\rangle $ under SLOCC.

From Theorem 1, we conclude the following corollary.

\textit{Corollary 1.} (Invariant AMs and GMs of eigenvalues and sizes of
JBs) If two states $|\psi \rangle $ and $|\psi ^{\prime }\rangle $\ of $4n$\
qubits\ are SLOCC equivalent, then the matrices $\Phi _{2^{2n+1}}(|\psi
^{\prime }\rangle )$ and $\Phi _{2^{2n+1}}(|\psi \rangle )$ have the same
AMs and GMs, and the JBs of $\Phi _{2^{2n+1}}(|\psi \rangle )$\ with the
eigenvalue $\lambda $\ and the JBs of $\Phi _{2^{2n+1}}(|\psi ^{\prime
}\rangle )$ with the eigenvalue$\sqrt{gh}\lambda $\ have the same sizes.

It is easy to derive the corollary from Theorem 1. The following is our
detailed argument. Eq. (\ref{P-S-3}) implies that $\Phi _{2^{2n+1}}(|\psi
^{\prime }\rangle )$ and $\Theta $ are similar. Therefore, $\Phi
_{2^{2n+1}}(|\psi ^{\prime }\rangle )$ and $\Theta $ have the same spectra
and SJNFs (ignoring the order of JBs).

(a). For the invariance of AMs

In light of Property 1 in Appendix B, if $\Phi _{2^{2n+1}}(|\psi \rangle )$
has the characteristic polynomial
\begin{eqnarray}
&&\det (\lambda I_{2^{2n+1}}-\Phi _{2^{2n+1}}(|\psi \rangle ))  \notag \\
&=&\lambda ^{2k}(\lambda \pm \lambda _{1})^{\ell _{1}}\cdots (\lambda \pm
\lambda _{s})^{\ell _{s}},  \label{am-1}
\end{eqnarray}%
where $\lambda _{i}\neq 0$ and $\lambda _{i}\neq \lambda _{j}$ when $i\neq j$%
, then $\Theta $ has the characteristic polynomial
\begin{eqnarray}
&&\det (\lambda I_{2^{2n+1}}-\Theta )  \notag \\
&=&\lambda ^{2k}(\lambda \pm \sqrt{gh}\lambda _{1})^{\ell _{1}}\cdots
(\lambda \pm \sqrt{gh}\lambda _{s})^{\ell _{s}}.  \label{am-2}
\end{eqnarray}%
Therefore, $\Phi _{2^{2n+1}}(|\psi ^{\prime }\rangle )$ and $\Phi
_{2^{2n+1}}(|\psi \rangle )$ have the same AMs.

(b). For the invariance of GMs

From Eqs. (\ref{am-1}, \ref{am-2}), if $\Phi _{2^{2n+1}}(|\psi \rangle )$
has the spectrum
\begin{equation*}
\{0^{\circledcirc 2k},(\pm \lambda _{1})^{\circledcirc \ell _{1}},\cdots
,(\pm \lambda _{s})^{\circledcirc \ell _{s}}\},
\end{equation*}%
where $\circledcirc \ell _{i}$\ stands for the AM $\ell _{i}$,\ then $\Theta
$ has the spectrum
\begin{equation*}
\{0^{\circledcirc 2k},(\pm \sqrt{gh}\lambda _{1})^{\circledcirc \ell
_{1}},\cdots ,(\pm \sqrt{gh}\lambda _{s})^{\circledcirc \ell _{s}}\}.
\end{equation*}

Let $\mu _{0}$ and $\mu _{0}^{\prime }$ be the GMs of the zero-eigenvalue of
$\Phi _{2^{2n+1}}(|\psi \rangle )$ and $\Theta $. In light of Property (2.2)
in Appendix B, $\mu _{0}=\mu _{0}^{\prime }$. In light of Property 7 in
Appendix C, the eigenvalues $\pm \lambda _{i}$ of $\Phi _{2^{2n+1}}(|\psi
\rangle )$ have the same GM, for example $\mu _{i}$. Similarly, the
eigenvalues $\pm \sqrt{gh}\lambda _{i}$ of $\Theta $ have the same GM, for
example $\mu _{i}^{\prime }$. In light of Property (2.1) in Appendix B, $\mu
_{i}=\mu _{i}^{\prime }$, $i=1,\cdots ,s$. Thus, $\Phi _{2^{2n+1}}(|\psi
\rangle )$ has the set of GMs $\{\mu _{0},\mu _{1},\cdots ,\mu _{s}\}$, and $%
\Phi _{2^{2n+1}}(|\psi ^{\prime }\rangle )$ has the set of GMs $\{\mu
_{0}^{\prime },\mu _{1}^{\prime },\cdots ,\mu _{s}^{\prime }\}$. Therefore, $%
\Phi _{2^{2n+1}}(|\psi ^{\prime }\rangle )$ and $\Phi _{2^{2n+1}}(|\psi
\rangle )$ have the same GMs.

(c). For the invariance of sizes of JBs

In light of Property (3) in Appendix B, $\Phi _{2^{2n+1}}(|\psi \rangle )$
has a JB with\ the size of $r$ corresponding to the eigenvalue $\lambda $ if
and only if $\Theta $ has a JB with\ the size of $r$ corresponding to the
eigenvalue $\sqrt{gh}\lambda $. The conclusion is also true for the
zero-eigenvalue. Therefore, the corresponding JBs of $\Phi _{2^{2n+1}}(|\psi
^{\prime }\rangle )$ and $\Phi _{2^{2n+1}}(|\psi \rangle )$\ have the same
sizes.

\section{Classification of spectra of matrices $\Phi _{2^{2n+1}}(|\protect%
\psi \rangle )$ and pure states of $4n$\ qubits via integer partitions of
the number $2^{2n}-k$}

In this paper, $\ell $ in $\lambda ^{\circledcirc \ell }$ indicates the AM
of the eigenvalue $\lambda $. If $\ell =1$, then we write $\lambda
^{\circledcirc 1}$ as $\lambda $. In this paper, let $P(i)$ be the number of
integer partitions of $i$. Specially, $P(0)=1$.

\subsection{For $4n$\ qubits via integer partitions}

By means of Property 1 in Appendix C, spectra of $\Phi _{2^{2n+1}}(|\psi
\rangle )$\ in Eq. (\ref{coef-1})\ are of the following form:%
\begin{equation}
\{0^{\circledcirc 2k},(\pm \lambda _{1})^{\circledcirc \ell _{1}},(\pm
\lambda _{2})^{\circledcirc \ell _{2}},\cdots ,(\pm \lambda
_{s})^{\circledcirc \ell _{s}}\},  \label{Jordan}
\end{equation}%
where $\lambda _{i}\neq 0$, $\lambda _{i}\neq \lambda _{j}$ when $i\neq j$, $%
0\leq k\leq 2^{2n}$, and $\ell _{i}$ is the AM of the eigenvalues $\pm
\lambda _{i}$. Clearly, all the AMs satisfy the equation
\begin{equation}
2(\ell _{1}+\ell _{2}+\cdots +\ell _{s})+2k=2^{2n+1},  \label{am-eq}
\end{equation}

Because the eigenvalues $\pm \lambda _{i}$ have the same AM $\ell _{i}$, for
the sake of simplicity and without loss of generality, we ignore $\pm $ in
Eq. (\ref{Jordan}) when calculating AMs below.

\subsubsection{A set of AMs is invariant under SLOCC and just an integer
partition of the number $2^{2n}-k$}

Let $\Xi $ be a set of AMs of eigenvalues in Eq. (\ref{Jordan}). Then,

\begin{equation}
\Xi =(2k;\ell _{1},\ell _{2},\cdots ,\ell _{s}),  \label{ams-1}
\end{equation}%
where $2k$ is the AM of the zero-eigenvalue while $\ell _{1}$, $\ell _{2}$, $%
\cdots $, and $\ell _{s}$ are the AMs\ of the different non-zero
eigenvalues. From Eq. (\ref{am-eq}), it is clear that $(\ell _{1},\ell
_{2},\cdots ,\ell _{s})$ is just an integer partition of the number $%
2^{2n}-k $, i.e.
\begin{equation}
\ell _{1}+\ell _{2}+\cdots +\ell _{s}=2^{2n}-k.  \label{parti-1}
\end{equation}%
In light of Corollary 1, $\Xi $\ is invariant under SLOCC.

\subsubsection{Classification via integer partitions of the number $2^{2n}-k$%
}

\paragraph{Spectra are partitioned into different types}

Next, we use $\Xi $\ to label spectra ignoring values of eigenvalues. For
example, we write $(0;1,1,2)$\ to label the spectrum $\{\lambda _{1},\lambda
_{2},\lambda _{3}{}^{\circledcirc 2}\}$ of a matrix $\Phi _{8}$, where $%
(1,1,2)$ is an integer partition of 4.

We define that spectra of matrices $\Phi _{2^{2n+1}}(|\psi \rangle )$\ in
Eq. (\ref{coef-1}) belong to the same type if the spectra have the same AMs,
i.e. the same $\Xi $ ignoring values of the eigenvalues. Thus, for spectra
of the same type, the sets of AMs of non-zero eigenvalues are the same
partition $(\ell _{1},\ell _{2},\cdots ,\ell _{s})$ of the number $2^{2n}-k$
for the same $k$.

For four qubits, we obtain 12 different types of spectra of $\Phi _{8}$
without considering permutations of qubits in Table I. In Table I, $SP$ is
short for a spectrum.

\begin{table}[tbph]
\caption{ 12 types of spectra of $\Phi _{8}$ for four qubits}%
\begin{ruledtabular}

\begin{tabular}{ccc}
Types &Spectra &$\Xi$\\

$SP_{1}$ & $\{(\pm \lambda _{1})^{\circledcirc 4}\}$ & (0;4)\\
$SP_{2}$ & $\{\pm \lambda _{1},(\pm \lambda_{2})^{\circledcirc 3}\} $ & (0; 1,3) \\
$SP_{3}$ & $\{\pm \lambda _{1},\pm \lambda _{2},(\pm \lambda _{3})^{\circledcirc 2}\}$ & (0;1,1,2) \\
$SP_{4}$& $\{(\pm \lambda _{1})^{\circledcirc 2},(\pm \lambda _{2})^{\circledcirc 2}\}$ & (0;2,2) \\
$SP_{5} $& $\{\pm \lambda _{1},\pm \lambda _{2},\pm \lambda _{3},\pm \lambda_{4}\} $ &  (0;1,1,1,1)\\
$SP_{6}$& $\{0^{\circledcirc 2},(\pm \lambda _{1})^{\circledcirc 3}\}$ &    (2;3) \\
$SP_{7}$ & $\{0^{\circledcirc 2},\pm \lambda _{1},(\pm \lambda _{2})^{\circledcirc 2}\}$ & (2;1,2)   \\
$SP_{8}$ & $\{0^{\circledcirc 2},\pm \lambda _{1},\pm \lambda _{2},\pm \lambda_{3}\}$  &   (2;1,1,1) \\
$SP_{9}$ & $\{0^{\circledcirc 4},(\pm \lambda _{1})^{\circledcirc 2}\} $ &   (4;2)  \\
$SP_{10} $&$\{0^{\circledcirc 4},\pm \lambda _{1},\pm \lambda _{2}\} $ &  (4;1,1) \\
$SP_{11}$ &$\{0^{\circledcirc 6},\pm \lambda _{1}\}$ &   (6;1)  \\
$SP_{12} $& $\{0^{\circledcirc 8}\}$ &(8; )

\end{tabular}

\end{ruledtabular}
\end{table}

\paragraph{Pure states are partitioned into different groups}

By letting pure states of $4n$ qubits with the same type of spectra of $\Phi
_{2^{2n+1}}(|\psi \rangle )$\ in Eq. (\ref{coef-1})\ belong to the same
group, then each group can be characterized with a set $\Xi $ of AMs. Thus,
SLOCC classification of $4n$ qubits is reduced to calculating integer
partitions of the number $2^{2n}-k$ for each $k$, where $0\leq k\leq 2^{2n}$.

One can know that for each partition $(\ell _{1},\ell _{2},\cdots ,\ell
_{s}) $ of $2^{2n}-k$, $(2k;\ell _{1},\ell _{2},\cdots ,\ell _{s})$
corresponds to a set of AMs of eigenvalues in Eq. (\ref{Jordan}). Different
partitions of $2^{2n}-k$ correspond to different types of spectra and
different groups of pure states. In light of Corollary 1, two pure states of
$4n$ qubits belonging to different groups are SLOCC inequivalent.

For the fixed $k$, from Eq. (\ref{parti-1}) there are $P(2^{2n}-k)$
different partitions of $2^{2n}-k$. For all $k$,\ a calculation yields $%
\sum_{i=0}^{2^{2n}}P(i)$ different partitions. From this, we can conclude
the following theorem.

\textit{Theorem 2. }Via partitions of $2^{2n}-k$, the matrices $\Phi
_{2^{2n+1}}(|\psi \rangle )$\ in Eq. (\ref{coef-1})\ have $%
\sum_{i=0}^{2^{2n}}P(i)$ different types of spectra and pure states of $4n$
qubits are classified into $\sum_{i=0}^{2^{2n}}P(i)$ different groups under
SLOCC.

\subsection{Classification of four qubits via integer partitions of $4-k$}

We first calculate partitions of $4-k$, where $0\leq k\leq 4$. For example,
for $k=1$, 3 ($=4-1$) can be partitioned in the three distinct ways: $3$, $%
1+2$, and $1+1+1$. Then, from the three partitions we obtain three sets of
AMs: (2;3), (2;1,2), and (2;1,1,1). For all $k$, there are 12 integer
partitions. So, there are 12 types of spectra of $\Phi _{8}$ and 12 groups
of pure states without considering permutations of qubits. Ref. Table I.

\subsection{Detect genuinely entangled states of $4n$ qubits via the
invariant $\Xi $}

For four qubits, 3 of 12 groups in the first column of Table III include
product states and we label the 3 groups with $\triangleleft $. Thus, other
9 groups are genuinely entangled, i.e. each state of the 9 groups is
genuinely entangled. For example, it is easy to check that $|\Upsilon
\rangle $ is genuinely entangled. Note that when calculating the invariant $%
\Xi $ for product states, we use the coefficient matrix $C_{12}(|\psi
\rangle )$.

For $4n$ qubits, if the spectrum of the matrix $\Phi _{2^{2n+1}}(|\psi
\rangle )$ does not belong to the types which include spectra of the
matrices $\Phi _{2^{2n+1}}(|\psi \rangle )$\ in Eq. (\ref{coef-1}) for
product states, then the state $|\psi \rangle $ is a genuinely entangled
state.

\section{Classification of SJNFs of matrices $\Phi _{2^{2n+1}}(|\protect\psi %
\rangle )$ and pure states of $4n$\ qubits\ via integer partitions of AMs}

In this paper, we write the direct sum $\underbrace{\text{ }J_{m}(\lambda
)\oplus \cdots \oplus J_{m}(\lambda )}_{j}$ as $J_{m}(\lambda )^{\oplus j}$
and the JB $J_{1}(a)$ as $a$.

\subsection{The relation between the set of sizes of JBs with the
zero-eigenvalue and the integer partition of the AM of the zero-eigenvalue}

Let $P^{\ast }(2k)$ be the number of different SJNFs with the spectrum $%
0^{\circledcirc 2k}$ by Properties 1, 3, and 5 in Appendix C, where $P^{\ast
}(0)=1$. To calculate $P^{\ast }(2k)$, we give the following definition.

Definition. If $m$ is partitioned into an even number of parts and in the
partition if a part is an even number then\ the number of its occurrences is
also even, then the partition is called a tri-even partition of $m$. For
example, the partition $2+2+3+1=8$ is a tri-even partition of 8 because 8 is
partitioned into four parts and \textquotedblleft 2\textquotedblright\
occurs twice.

One can check that in light of Properties 1, 3, and 5 in Appendix C, the set
of sizes of JBs with the zero-eigenvalue must be a tri-even partition of $2k$
for the spectrum $0^{\circledcirc 2k}$. Conversely, the JBs with the
zero-eigenvalue, of which the set of sizes\ is a tri-even partition of $2k$
for the spectrum $0^{\circledcirc 2k}$, must satisfy Properties 1, 3, and 5
in Appendix C.

For the spectrum $0^{\circledcirc 2^{2n+1}}$, we do not consider the integer
partition $(\underbrace{1,\cdots ,1}_{2^{2n+1}})$ of $2^{2n+1}$ which
implies that the corresponding SJNF is the zero matrix, then $\Phi
_{2^{2n+1}}(|\psi \rangle )$\ $=0$, and then all the coefficients of the
corresponding state vanish.

Let $\widetilde{2k}$ be a set of all the tri-even partition of $2k$, where $%
\tilde{0}=\phi $, which is the empty set. A simple calculation yields that $%
P^{\ast }(2)=1$, $P^{\ast }(4)=3$, $P^{\ast }(6)=5$, and $P^{\ast }(8)=10$.

\subsection{Classification for$\ 4n$\ qubits via integer partitions of AMs}

In light of Property 3 in Appendix C, the number of JBs corresponding to the
zero-eigenvalue of $\Phi _{2^{2n+1}}(|\psi \rangle )$\ in Eq. (\ref{coef-1}%
)\ is even. In light of Property 7 in Appendix C, the numbers of JBs
corresponding to the non-zero eigenvalues $\pm \lambda $ of $\Phi
_{2^{2n+1}}(|\psi \rangle )$\ in Eq. (\ref{coef-1}) are the same. In light
of Property 8 in Appendix C, the corresponding JBs with the non-zero
eigenvalues $\pm \lambda $ have the same size. Thus, SJNFs of\ $\Phi
_{2^{2n+1}}(|\psi \rangle )$\ in Eq. (\ref{coef-1})\ with the spectrum in
Eq. (\ref{Jordan}) are of the following form:%
\begin{eqnarray}
&&J_{\tau _{1}}(0)\cdots J_{\tau _{2m}}(0)J_{\alpha _{1}}(\pm \lambda
_{1})\cdots J_{\alpha _{l_{1}}}(\pm \lambda _{1})  \notag \\
&&J_{\beta _{1}}(\pm \lambda _{2})\cdots J_{\beta _{l_{2}}}(\pm \lambda
_{2})\cdots J_{\gamma _{1}}(\pm \lambda _{s})\cdots J_{\gamma _{l_{s}}}(\pm
\lambda _{s}).  \notag \\
&&  \label{SJNF-1}
\end{eqnarray}

For four qubits, there are 43 different SJNFs of $\Phi _{8}$ in Table II
without considering permutations of qubits. Note that in Table II, $\lambda
_{i}$ are the eigenvalues of $\Phi _{8}$, where $\lambda _{i}\neq 0$ and $%
\lambda _{i}\neq \lambda _{j}$ when $i\neq j$.

Note that Table II$\ $does not include the SJNFs: $\pm \lambda
_{1}J_{2}(0)J_{4}(0)$, $J_{2}(0)J_{6}(0)$ or $00J_{2}(0)J_{4}(0)$. This is
because these SJNFs do not satisfy Property 5.1 in Appendix C.\

\begin{table}[tbph]
\caption{ 43 different types of SJNFs of $\Phi _{8}$ for four qubits
corresponding to the spectra $SP_{1}-SP_{12}$ in Table I}
\label{tab2}%
\begin{ruledtabular}
\begin{tabular}{ccc}

$SP$ & $\text{SJNF}$ & $\text{SJNF}$ \\ \hline
$SP_{1}$ & $\text{ }J_{4}(\pm \lambda _{1})$
& $\text{ }J_{2}(\pm \lambda_{1})J_{2}(\pm \lambda _{1})$ \\
& $J_{3}(\pm \lambda _{1})\pm \lambda _{1}$
& $(\pm \lambda _{1})^{\oplus2}J_{2}(\pm \lambda _{1})$ \\
& $(\pm \lambda _{1})^{\oplus 4}\ \ddag_{1}$ &  \\  \hline
$SP_{2}$ & $\pm \lambda _{1}J_{3}(\pm \lambda _{2})$
 & $\pm \lambda _{1}\pm\lambda _{2}J_{2}(\pm \lambda _{2})$ \\
& $\pm \lambda _{1}(\pm \lambda _{2})^{\oplus 3}$ &  \\ \hline
$SP_{3}$ & $\pm \lambda _{1}\pm \lambda _{2}J_{2}(\pm \lambda _{3})$
 & $\pm\lambda _{1}\pm \lambda _{2}(\pm \lambda _{3})^{\oplus 2}$ \\ \hline
$SP_{4}$ & $(\pm \lambda _{1})^{\oplus 2}(\pm \lambda _{2})^{\oplus 2}$
& $ (\pm\lambda _{1})^{\oplus 2}J_{2}(\pm \lambda _{2})$ \\
& $J_{2}(\pm \lambda _{1})J_{2}(\pm \lambda _{2})$ &  \\ \hline
$SP_{5}$ & $\pm \lambda _{1}\pm \lambda _{2}\pm \lambda _{3}\pm \lambda _{4}$ &
\\  \hline
$SP_{6}$ & $00J_{3}(\pm \lambda _{1})$
&$ 00J_{2}(\pm \lambda _{1})\pm \lambda_{1}$ \\
& $00(\pm \lambda _{1})^{\oplus 3}$ &  \\ \hline
$SP_{7} $& $00\pm \lambda _{1}J_{2}(\pm \lambda _{2})$
& $00\pm \lambda_{1}J_{1}(\pm \lambda _{2})^{{\tiny \oplus }2}$ \\ \hline
$SP_{8} $& $00\pm \lambda _{1}\pm \lambda _{2}\pm \lambda _{3}$ &  \\ \hline
$SP_{9}$ & $J_{2}(0)^{\oplus 2}J_{2}(\pm \lambda _{1})$
& $J_{2}(0)^{\oplus2}(\pm \lambda _{1})^{\oplus 2}$ \\
& $J_{3}(0)0J_{2}(\pm \lambda _{1})$ & $J_{3}(0)0(\pm \lambda _{1})^{\oplus 2}$
\\
& $0^{{\tiny \oplus }4}J_{2}(\pm \lambda _{1})$
& $0^{\oplus 4}\pm \lambda_{1}\pm \lambda _{1}$ \\ \hline
$SP_{10}$ & $J_{2}(0)^{\oplus 2}\pm \lambda _{1}\pm \lambda _{2} $
&$J_{3}(0)0\pm\lambda _{1}\pm \lambda _{2}$ \\
& $0^{\oplus 4}\pm \lambda _{1}\pm \lambda _{2}$ &  \\ \hline
$SP_{11}$ & $0J_{5}(0)\pm \lambda _{1}$ & $J_{3}(0)^{\oplus 2}\pm \lambda _{1}$ \\
& $J_{3}(0)0^{\oplus 3}\pm \lambda _{1}$ & $J_{2}(0)^{\oplus 2}00\pm \lambda_{1} $\\
&$ 0^{\oplus 6}\pm \lambda _{1}\ \ddag_{2}$ &\\ \hline

$SP_{12}$ &$ J_{7}(0)0$ &$ J_{5}(0)J_{3}(0)$ \\
& $J_{4}(0)^{\oplus 2}$ &  \\
 & $J_{2}(0)^{\oplus 4} \ \ddag_{3}$ & $J_{3}(0)^{\oplus 2}0^{\oplus 2}\ \ddag_{4}$ \\
& $J_{2}(0)^{\oplus 2}J_{3}(0)0\ \ddag_{5}$ & $J_{5}(0)0^{\oplus 3}$ \\
 & $J_{2}(0)^{\oplus 2}0^{\oplus 4}\ \ddag_{6}$ & $J_{3}(0)0^{\oplus 5}\ \ddag_{7}$\\

\footnotetext [0]{$\ddag_{1}$ includes the product states
$|$EPR $\rangle _{13}|$EPR $\rangle _{24}$ and
$|$EPR $\rangle _{14}|$ EPR $\rangle_{23}$.
\newline
$\ddag{2}$ includes the product state
$|$ EPR $\rangle _{12}|$ EPR $\rangle _{34}$.
\newline
$\ddag_{3}$ includes the product states $|00\rangle
_{13}|$EPR$\rangle _{24}$, $|00\rangle _{14}|$EPR$\rangle _{23}$,
$|00\rangle _{23}|$EPR$\rangle _{14}$, and $|00\rangle _{24}|$EPR$\rangle _{13}$.
\newline
$\ddag_{4} $ includes the product states $|0\rangle _{i}|$GHZ$\rangle _{jkl}$.
\newline
$\ddag_{5}$ includes the product states $|0\rangle _{i}|$W$\rangle _{jkl}$.
\newline
$\ddag{6}$ includes the full separate state $|0000\rangle $.
\newline
$\ddag{7}$ includes the product states $|00\rangle _{12}|$EPR$\rangle
_{34}$ and $|00\rangle _{34}|$EPR$\rangle _{12}$. }
\end{tabular}
\end{ruledtabular}
\end{table}

Note that each pair of JBs like $J_{\alpha _{1}}(\pm \lambda _{1})$ in Eq. (%
\ref{SJNF-1}) have the same size. For the sake of simplicity and without
loss of generality, we ignore $\pm $ in JBs in Eq. (\ref{SJNF-1}) when
calculating sizes of JBs below. For example, for the SJNF $J_{2}(\pm \lambda
_{1})J_{2}(\pm \lambda _{2})$\ of $\Phi _{8}$, we only consider the sizes of
the JBs $J_{2}(\lambda _{1})$ and $J_{2}(\lambda _{2})$ ignoring the sizes
of the JBs $J_{2}(-\lambda _{1})$ and $J_{2}(-\lambda _{2})$.

\subsubsection{A collection of sets of sizes of JBs with different
eigenvalues is invariant under SLOCC and just a list of partitions of AMs}

In Eq. (\ref{SJNF-1}), let $\tau $ be a set of sizes of JBs with the
zero-eigenvalue and $\pi _{1}$ (resp. $\pi _{2}$,$\cdots $, $\pi _{s}$) be a
set of sizes of JBs with the eigenvalue $\lambda _{1}$ (resp. $\lambda _{2}$,%
$\cdots $,$\lambda _{s}$). From Eq. (\ref{SJNF-1}), we obtain
\begin{eqnarray*}
\tau &=&(\tau _{1},\cdots ,\tau _{2m}), \\
\pi _{1} &=&(\alpha _{1},\cdots ,\alpha _{l_{1}}), \\
\pi _{2} &=&(\beta _{1},\cdots ,\beta _{l_{2}}), \\
&&\vdots \\
\pi _{s} &=&(\gamma _{1},\cdots ,\gamma _{l_{s}}).
\end{eqnarray*}%
Let
\begin{equation}
\vartheta =\{\tau ;\pi _{1},\pi _{2},\cdots ,\pi _{s}\}.  \label{list-1-}
\end{equation}%
In light of Corollary 1, $\vartheta $ is invariant under SLOCC.

Clearly, each SJNF\ can be described by the $\vartheta $. For example, for
the SJNF $J_{2}(\lambda _{1})J_{2}(\lambda _{2})$, $\tau =\phi $ and $%
\vartheta =\{\phi ;(2),(2)\}$. For the SJNF $J_{2}(\lambda _{1})\lambda
_{2}\lambda _{2}$, $\vartheta =\{\phi ;(2),(1,1)\}$. We call $\vartheta $
the label of the SJNF.

From the above discussion, $\tau $ is just a tri-even partition of$\ 2k$
(here, $2k$ is the AM\ of the zero-eigenvalue), and $\pi _{1}$ (resp. $\pi
_{2}\cdots ,\pi _{s}$) is just a partition of $\ell _{1}$ (resp. $\ell _{2}$,%
$\cdots $, $\ell _{s}$) which is the AM\ of the eigenvalue $\lambda _{1}$
(resp. $\lambda _{2}$,$\cdots $, $\lambda _{s}$). Ref. Eq. (\ref{Jordan}).

In this paper, let $\overline{l}$ stand for a set of all the integer
partitions of $l$. For example, $\overline{2}=\{(2),(1,1)\}$ and $\overline{3%
}=\{(3)$, $(2,1)$, $(1,1,1)\}$. Thus, $\tau \in \widetilde{2k}$ , $\pi
_{i}\in \overline{\ell _{i}}$, $i=1$,$\cdots $, $s$. Clearly, $\vartheta $
is also a list of partitions of AMs (ref. Eq. (\ref{Jordan})), and thus each
SJNF corresponds to a list of partitions of AMs ignoring values of
eigenvalues.

\subsubsection{Classification of SJNFs and pure states\ via integer
partitions of AMs}

\paragraph{SJNFs are partitioned into different types}

For example, for the SJNFs $J_{2}(\lambda _{1})\lambda _{2}\lambda _{2}$ and
$\lambda _{1}\lambda _{1}J_{2}(\lambda _{2})$, one can see that one of the
two SJNFs can be obtained from the other one by renaming $\lambda _{1}$ as $%
\lambda _{2}$ and $\lambda _{2}$ as $\lambda _{1}$ simultaneously. Here, we
consider that these two SJNFs possess the same type. Note that for the SJNF $%
J_{2}(\lambda _{1})\lambda _{2}\lambda _{2}$, the labels is $\vartheta
=\{\phi ;(2),(1,1)\}$ and for the SJNF $\lambda _{1}\lambda
_{1}J_{2}(\lambda _{2})$, the label $\vartheta ^{\prime }=\{\phi
;(1,1),(2)\} $. Here, we also consider that $\vartheta =\vartheta ^{\prime }$
ignoring the order of $(2)$ and $(1,1)$.

Generally, for two labels
\begin{equation*}
\vartheta =\{\tau ;\pi _{1},\pi _{2},\cdots ,\pi _{s}\}.
\end{equation*}%
and
\begin{equation*}
\vartheta ^{\prime }=\{\tau ^{\prime };\pi _{1}^{\prime },\pi _{2}^{\prime
},\cdots ,\pi _{s}^{\prime }\},
\end{equation*}%
we define that $\vartheta =\vartheta ^{\prime }$ if and only if
\begin{equation*}
\tau =\tau ^{\prime }
\end{equation*}%
and
\begin{equation*}
\{\pi _{1},\pi _{2}\cdots ,\pi _{s}\}=\{\pi _{1}^{\prime },\pi _{2}^{\prime
},\cdots ,\pi _{s}^{\prime }\}
\end{equation*}%
ignoring the order of $\pi _{1},\pi _{2},\cdots ,$ and $\pi _{s}$ and the
order of $\pi _{1}^{\prime },\pi _{2}^{\prime },\cdots ,$ and $\pi
_{s}^{\prime }$.

We can next define that two SJNFs possess the same type if and only if their
labels are equal.

For four qubits, we obtain 43 types of SJNFs. Ref. Table II and the second
and third columns of Table III.

\paragraph{Pure states are partitioned into different families}

By letting states with the same type of SJNFs of $\Phi _{2^{2n+1}}(|\psi
\rangle )$\ in Eq. (\ref{coef-1})\ belong to the same family, then each
family can be described with an invariant $\vartheta =\{\tau ;\pi _{1},\pi
_{2}\cdots ,\pi _{s}\}$. Thus, SLOCC classification of $4n$ qubits\ is
reduced to calculating integer partitions of AMs.

We next explain how to calculate all the integer partitions of AMs. We first
calculate partitions of $2^{2n}-k$ for each $k$, where $0\leq k\leq 2^{2n}$.
Then, for each partition $(\ell _{1},\ell _{2},\cdots ,\ell _{s})$ of $%
2^{2n}-k$, we calculate partitions of $\ell _{i}$ ($i=1,\cdots ,s$) and
tri-even partitions of $2k$. Conversely, let $\tau \in \widetilde{2k}$, $\pi
_{i}\in \overline{\ell _{i}}$, $i=1$,$\cdots $, $s$. Then, the list of
partitions$\ \vartheta $ ($=\{\tau ;\pi _{1},\pi _{2}\cdots ,\pi _{s}\}$)
corresponds to a collection of sets of sizes of JBs of $\Phi
_{2^{2n+1}}(|\psi \rangle )$\ in Eq. (\ref{coef-1}) for $4n$ qubits.

One can see that different $\vartheta $ correspond to different types of
SJNFs of $\Phi _{2^{2n+1}}(|\psi \rangle )$\ in Eq. (\ref{coef-1})\ and
different families of pure states. In light of Corollary 1, two states
belonging to different families are SLOCC inequivalent.

In Appendix D, a calculation shows that there are $\eta -1$\ different lists
of partitions of AMs, where $\eta $ is defined in Eq. (\ref{set-p-2}).\
Then, we can conclude the following theorem.

\textit{Theorem 3.} Via partitions of AMs, i.e. via partitions of $\ell _{i}$
($i=1,\cdots ,s$) and tri-even partitions of $2k$ in each partition $(\ell
_{1},\ell _{2},\cdots ,\ell _{s})$ of $2^{2n}-k$, where $0\leq k\leq 2^{2n}$%
, $\Phi _{2^{2n+1}}(|\psi \rangle )$\ in Eq. (\ref{coef-1}) has $\eta -1$
different types of SJNFs\ \ and then, pure states of $4n$ qubits are
classified into $\eta -1$ different families.

\subsection{Classification of four qubits}

We first calculate partitions of $4-k$ for each $k$, where $0\leq k\leq 4$.
For all $k$, there are 12 partitions. Then, for each partition $(\ell
_{1},\ell _{2},\cdots ,\ell _{s})$ of $4-k$, we calculate partitions of $%
\ell _{i}$ ($i=1,\cdots ,s$) and tri-even partitions of $2k$. For example,
let $k=0$,\ for the partition $(2,2)$ of 4, we calculate partitions of 2,
then we obtain three different lists of partitions: $\{\phi ;(1,1),(1,1)\}$,
$\{\phi ;(1,1),(2)\}$, and $\{\phi ;(2),(2)\}$.

For four qubits, in total\ there are 43 different lists of partitions. Thus,
we obtain 43 different types of SJNFs and 43 SLOCC\ inequivalent families of
pure states without considering permutations of qubits. Ref. the second and
third columns of Table III.

Furthermore, for each type of SJNFs, we can give a state of four qubits for
which $\Phi _{8}$ has the corresponding type.

\begin{table}[tbph]
\caption{ SLOCC classification of four qubits }
\label{tab1}%
\begin{ruledtabular}
\begin{tabular}{ccc}
 $\Xi$ &   $\vartheta $&   $\vartheta $\\ \hline
 (0;4) $\lhd _{1}$& \{$\phi ;$(4)\} & \{$\phi ;$(2,2)\} \\
 & \{$\phi ;$(1,1,2)\} & \{$\phi ;$(1,1,1,1)\} $\ddagger _{1}$ \\
 & \{$\phi ;$(3,1)\} &  \\ \hline
(0; 1,3) & \{$\phi ;$(1),(3)\} & \{$\phi ;$(1),(1,2)\} \\
  & \{$\phi ;$(1),(1,1,1)\} &  \\ \hline
(0;1,1,2) & \{$\phi ;$(1),(1),(2)\} & \{$\phi ;$(1),(1),(1,1)\} \\ \hline
 (0;2,2) & \{$\phi ;$(1,1),(1,1)\} & \{$\phi ;$(1,1),(2)\} \\
 & \{$\phi ;$(2),(2)\} &  \\ \hline
 (0;1,1,1,1) & \{$\phi ;$(1),(1),(1),(1)\} &  \\ \hline
 (2;3) & \{\underline{(1,1)};(3)\} & \{\underline{(1,1)};(2,1)\} \\
 & \{\underline{(1,1)};(1,1,1)\} &  \\ \hline
 (2;1,2) & \{\underline{(1,1)};(1),(2)\} & \{\underline{(1,1)};(1),(1,1)\}
\\ \hline
(2;1,1,1) & \{\underline{(1,1)};(1),(1),(1)\} &  \\ \hline
 (4;2) & \{\underline{(2,2)};(2)\} & \{\underline{(2,2)};(1,1)\} \\
 & \{\underline{(3,1)};(2)\} & \{\underline{(3,1)};(1,1)\} \\
 & \{\underline{(1,1,1,1)};(2)\} & \{\underline{(1,1,1,1)};(1,1)\} \\ \hline
 (4;1,1) & \{\underline{(2,2)};(1),(1)\} & \{\underline{(3,1)};(1),(1)\} \\
 & \{\underline{(1,1,1,1)};(1),(1)\} &  \\ \hline
 (6;1)  $\lhd _{2}$& \{\underline{(1,5)};(1)\} & \{\underline{(3,3)};(1)\} \\
 & \{\underline{(2,2,1,1)};(1)\} & \{\underline{(1,1,1,1,1,1)};(1)\} $%
\ddagger _{2}$ \\
  & \{\underline{(3,1,1,1)};(1)\} &  \\ \hline

 (8; )$\lhd _{3}$ &  \{\underline{(7,1)};\} & \{\underline{(5,3)};\} \\
 &  \{\underline{(4,4)};\} & \{\underline{(2,2,2,2)};\} $\ \ddag_{3}$\\
 & \{\underline{(3,3,1,1)};\} $\ \ddag_{4}$& \{\underline{(2,2,3,1)};\}$\ \ddag_{5}$  \\
 & \{\underline{(5,1,1,1)};\}&\{\underline{(2,2,1,1,1,1)};\} $\ \ddag_{6}$\\
 &\{\underline{(3,1,1,1,1,1)};\}$\ \ddag_{7}$ & \\

\footnotetext [0]
{$\phi $ is the empty set. $\underline{(\cdots ) }$
 is the set of sizes of JBs with the zero-eigenvalue.
 \newline
$\ddag_{1}$ includes
$|$EPR $\rangle _{13}|$EPR $\rangle _{24}$ and
$|$EPR $\rangle _{14}|$ EPR $\rangle_{23}$.
\newline
$\ddag{2}$ includes
$|$ EPR $\rangle _{12}|$ EPR $\rangle _{34}$.
\newline
$\ddag_{3}$ includes  $|00\rangle
_{13}|$EPR$\rangle _{24}$, $|00\rangle _{14}|$EPR$\rangle _{23}$,
$|00\rangle _{23}|$EPR$\rangle _{14}$, and $|00\rangle _{24}|$EPR$\rangle _{13}$.
$\ddag_{4} $ includes  $|0\rangle _{i}|$GHZ$\rangle _{jkl}$.
$\ddag_{5}$ includes  $|0\rangle _{i}|$W$\rangle _{jkl}$.
$\ddag{6}$ includes $|0000\rangle $.
\newline
$\ddag{7}$ includes  $|00\rangle _{12}|$EPR$\rangle
_{34}$ and $|00\rangle _{34}|$EPR$\rangle _{12}$.
\newline
$\lhd _{1}$ includes  $|$EPR$%
\rangle _{13}|$EPR$\rangle _{24}$ and $|$EPR$\rangle _{14}|$EPR$\rangle _{23}$.
\newline
$\lhd _{2}$ includes
 $|$EPR$\rangle _{12}|$EPR$\rangle _{34}$.
\newline
$\lhd _{3}$ includes  $|0\rangle _{i}|$W$%
\rangle _{jkl}$, $|0\rangle _{i}|$GHZ$\rangle _{jkl}$, $|0\rangle
_{i}|0\rangle _{j}|$EPR$\rangle _{kl}$, and $|0000\rangle $,
where $|$GHZ$\rangle _{jkl}$ is a 3-qubit GHZ state,
 $|$W$\rangle _{jkl}$ is a 3-qiubit W state,
 and $|$EPR$\rangle _{kl}$ is a 2-qubit EPR state.
 }
\end{tabular}
\end{ruledtabular}
\end{table}

Note that Table III does not include the following $\vartheta $: $\{%
\underline{(2,4)};(1)\}$, $\{\underline{(2,6)}\};\}$, $\{\underline{%
(1,1,2,4);}\}$. This is because the corresponding SJNFs do not satisfy
Property 5.1 in Appendix C.\

\subsection{Detect genuinely entangled states of $4n$ qubits via the
invariant $\protect\vartheta $}

For four qubits, 7 of 43 families (ref. the second and third columns of
Table III) include product states and we label the 7 families with $\ddagger
$ in Table III. Thus, other 36 families are genuinely entangled, i.e. each
state of the 36 families is genuinely entangled. For example, it is easy to
check that $|\Upsilon \rangle $ is genuinely entangled. Note that when
calculating the invariant $\vartheta $ for product states we use the
coefficient matrix $C_{12}(|\psi \rangle )$.

One can see that only four families $L_{ab_{3}}^{\ast }$, $L_{a_{4}}$, $%
L_{0_{5\oplus 3}}$, and $L_{0_{7\oplus 1}}$ of Verstraete et al.'s nine
families\ are genuinely entangled, where $L_{ab_{3}}^{\ast }$ is obtained by
replacing the last two $+$ signs of $L_{ab_{3}}$ with $-$ signs \cite%
{Chterental}.

For $4n$ qubits, if the SJNF of the matrix $\Phi _{2^{2n+1}}(|\psi \rangle )$
does not belong to the types which include SJNFs of matrices $\Phi
_{2^{2n+1}}(|\psi \rangle )$\ in Eq. (\ref{coef-1})\ for product states,
then the state $|\psi \rangle $ is a genuinely entangled state.

\section{Comparison to Verstraete et al.'s nine families}

Via the complex SVD, Verstraete et al. partitioned pure states of four
qubits into nine families: $G_{abcd}$, $L_{abc_{2}}$, $L_{a_{2}b_{2}}$, $%
L_{ab_{3}}$, $L_{a_{4}}$, $L_{a_{2}0_{3\oplus 1}}$, $L_{0_{5\oplus 3}}$, $%
L_{0_{7\oplus 1}}$, $L_{0_{3\oplus 1}0_{3\oplus 1}}$ up to permutations of
the qubits under determinant 1 SLOCC \cite{Verstraete}.

In this paper, we show that if two pure states of $4n$ qubits are SLOCC
equivalent, then the spectra and SJNFs of their matrices $\Phi
_{2^{2n+1}}(|\psi \rangle )$\ in Eq. (\ref{coef-1}) are proportional. It
means the invariance of AMs and GMs and the sizes of JBs. Via integer
partitions of $2^{2n}-k$, we can partition pure states of $4n$ qubits into $%
\sum_{i=0}^{2^{2n}}P(i)$ different groups under SLOCC\ without considering
permutations of qubits. Specially, pure states of four qubits are
partitioned into 12 types. Via integer partitions of AMs, we can partition
pure states of $4n$ qubits into $\eta -1$ ($\eta $ is defined in Eq. (\ref%
{set-p-2})) different families under SLOCC\ without considering permutations
of qubits. Specially, pure states of four qubits into are partitioned into
43 families.

Chterental and Djokovi\'{c} pointed out an error in Verstraete et al.'s nine
families by indicating that the family$\ L_{ab_{3}}$ is SLOCC\ equivalent to
the subfamily $L_{abc_{2}}(a=c)$ of the family $L_{abc_{2}}$ \cite%
{Chterental}. The statement was corrected in \cite{LDFPRA15}, where it was
deduced\ that when $a\neq 0$, the family $L_{ab_{3}}$ is SLOCC\ equivalent
to the subfamily $L_{abc_{2}}(a=c)$ of the family $L_{abc_{2}}$ while $a=0$,
$L_{ab_{3}}$ and $L_{abc_{2}}(a=c)$ are SLOCC inequivalent. In light of
Theorem 1, we can also show that $L_{ab_{3}}(a=0)$ and $L_{abc_{2}}(a=c=0)$
are SLOCC inequivalent because the matrices $\Phi _{8}$ have SJNFs $%
J_{3}(0)J_{3}(0)\pm b$ and $J_{2}(0)J_{2}(0)\pm b00$ for $L_{ab_{3}}(a=0)$
and $L_{abc_{2}}(a=c=0)$, respectively.

For the completeness of Verstraete et al.'s nine families, Chterental and
Djokovi\'{c} changed the family $L_{ab_{3}}$ as the family $L_{ab_{3}}^{\ast
}$ defined above. A calculation yields that the states $L_{ab_{3}}(a=b=0)$, $%
L_{ab_{3}}^{\ast }(a=b=0)$, and $|0\rangle (|000\rangle +|111\rangle )$,
which is the representative state of the family $L_{0_{3\oplus 1}0_{3\oplus
1}}$, have the same Jordan block structure $J_{3}(0)J_{3}(0)00$ though the
states $L_{ab_{3}}(a=b=0)$ and $|0\rangle (|000\rangle +|111\rangle )$, and
the states $L_{ab_{3}}^{\ast }(a=b=0)$ and $|0\rangle (|000\rangle
+|111\rangle )$ are SLOCC\ inequivalent, respectively. Note that $%
L_{ab_{3}}(a=b=0)$ is SLOCC equivalent to $L_{ab_{3}}^{\ast }(a=b=0)$ \cite%
{LDFPRA12}.

Recall that a family is defined as having Jordan and degenerated Jordan
blocks of specific dimension (see the proof of Theorem 2 on page 3 of \cite%
{Verstraete}). So, via the definition for the families, the states $%
L_{ab_{3}}(a=b=0)$ and $|0\rangle (|000\rangle +|111\rangle )$ should belong
to the same family, and the states $L_{ab_{3}}^{\ast }(a=b=0)$ and $%
|0\rangle (|000\rangle +|111\rangle )$ should belong to the same family.
Unfortunately, they are partitioned into different families. Clearly, the
definition for the families and the representative states are not consistent
and pure states of four qubits are partitioned into the nine families
incompletely. These errors are avoided in this paper. In this paper, the
three states $L_{ab_{3}}^{\ast }(a=b=0)$, $L_{ab_{3}}(a=b=0)$, and $%
|0\rangle (|000\rangle +|111\rangle $ are included in one family.

\section{Summary}

In this paper, we show that algebraic and geometric multiplicities of
eigenvalues and sizes of JBs of $\Phi _{2^{2n+1}}(|\psi \rangle )$\ in Eq. (%
\ref{coef-1}) are invariant under SLOCC. Thus, we have invariants $\Xi
=(2k;\ell _{1},\ell _{2},\cdots ,\ell _{s})$ and $\vartheta =\{\tau ;\pi
_{1},\pi _{2}\cdots ,\pi _{s}\}$, where $\tau \in \widetilde{2k}$, $\pi
_{i}\in \overline{\ell _{i}}$ ( $i=1,\cdots ,s$), $2k$ is the AM of the
zero-eigenvalue, $\ell _{1},\ell _{2},\cdots ,\ell _{s}$ are the AMs of the
non-zero eigenvalues, $\overline{\ell _{i}}$ is a set of all the integer
partitions of $\ell _{i}$, and $(\ell _{1},\ell _{2},\cdots ,\ell _{s})$ is
just a partition of $2^{2n}-k$. Note that $\vartheta $ is also a collection
of sets of sizes of JBs.

For $4n$\ qubits, for all $k$\ there are $\sum_{i=0}^{2^{2n}}P(i)$ different
partitions of $2^{2n}-k$. For four qubits, for all $k$ there are 12
partitions of $4-k$. Ref. the first column of Table III. Thus, for $4n$\
qubits, we obtain $\sum_{i=0}^{2^{2n}}P(i)$ different types of spectra and
then classify pure states of $4n$ qubits into $\sum_{i=0}^{2^{2n}}P(i)$
different groups. Specially, pure states of four (eight) qubits are
partitioned into 12 (915) groups.

Furthermore, for each partition $(\ell _{1},\ell _{2},\cdots ,\ell _{s})$\
of $2^{2n}-k$, by calculating partitions of $\ell _{i}$ and tri-even
partitions of $2k$ we can obtain $\eta -1$\ different lists of partitions,
then $\eta -1$ different types of SJNFs of $\Phi _{2^{2n+1}}(|\psi \rangle )$%
\ in Eq. (\ref{coef-1})\ and $\eta -1$ different families of pure states of $%
4n$ qubits. Specially, for four qubits, we obtain 43 families. Ref. the
second and third columns of Table III. We show that 9 of 12 groups and 36 of
43 families are genuinely entangled.

We also show that if spectra or SJNFs of\ two matrices $\Phi
_{2^{2n+1}}(|\psi \rangle )$\ in Eq. (\ref{coef-1}) associated with two $4n$%
-qubit pure states are not proportional, then the two states are SLOCC
inequivalent.

Acknowledgement---This work was supported by Tsinghua National Laboratory
for Information Science and Technology.

\section*{Appendix A A calculation of $Q_{i}Q_{i}^{t}$\ }


\setcounter{equation}{0} \renewcommand{\theequation}{A\arabic{equation}}

We calculate $Q_{i}Q_{i}^{t}$, $i=1,2$, as follows. First we show that
\begin{equation}
U^{+}U^{\ast }=\upsilon ^{\otimes 2n},  \label{eq-1}
\end{equation}%
where $U^{\ast }$ is a complex conjugate of $U$. Eq. (\ref{eq-1}) holds from
$T^{+}T^{\ast }=\upsilon \otimes \upsilon $ and $U^{+}U^{\ast }=T^{+}T^{\ast
}\otimes \cdots \otimes T^{+}T^{\ast }$, where $\upsilon =\left(
\begin{array}{cc}
0 & 1 \\
-1 & 0%
\end{array}%
\right) $. Then, a calculation yields
\begin{equation}
Q_{1}Q_{1}^{t}=U\Delta _{1}U^{+}U^{\ast }\Delta _{1}^{t}U^{t}.
\end{equation}%
Via Eq. (\ref{eq-1}),%
\begin{equation}
Q_{1}Q_{1}^{t}=U\Delta _{1}\upsilon ^{\otimes 2n}\Delta _{1}^{t}U^{t}.
\end{equation}

Using the definitions for $U$ and $\Delta _{1}$, a straightforward
calculation derives
\begin{eqnarray}
&&Q_{1}Q_{1}^{t}=T^{\otimes n}(\otimes _{i=1}^{2n}\mathcal{A}%
_{q_{i}})\upsilon ^{\otimes 2n}(\otimes _{i=1}^{2n}\mathcal{A}%
_{q_{i}}^{t})(T^{t})^{\otimes n}  \notag \\
&=&[T(\mathcal{A}_{q_{1}}\otimes \mathcal{A}_{q_{2}})\upsilon ^{\otimes 2}(%
\mathcal{A}_{q_{1}}^{t}\otimes \mathcal{A}_{q_{2}}^{t})T^{t}]\otimes \cdots
\otimes  \notag \\
&&[T(A_{q_{2n-1}}\otimes A_{q_{2n}})\upsilon ^{\otimes
2}(A_{q_{2n-1}}^{t}\otimes A_{q_{2n}}^{t})T^{t}].  \label{eq-2}
\end{eqnarray}%
Next we reduce Eq. (\ref{eq-2}). It is easy to test%
\begin{equation}
\mathcal{A}_{i}\upsilon \mathcal{A}_{i}^{t}=(\det \mathcal{A}_{i})\upsilon
\end{equation}%
and

\begin{eqnarray}
&&(\mathcal{A}_{i}\otimes \mathcal{A}_{j})\upsilon ^{\otimes 2}(\mathcal{A}%
_{i}^{t}\otimes \mathcal{A}_{j}^{t})  \notag \\
&=&\mathcal{A}_{i}\upsilon \mathcal{A}_{i}^{t}\otimes \mathcal{A}%
_{j}\upsilon \mathcal{A}_{j}^{t} \\
&=&(\det \mathcal{A}_{i})\upsilon \otimes (\det \mathcal{A}_{j})\upsilon \\
&=&(\det \mathcal{A}_{i})(\det \mathcal{A}_{j})\upsilon ^{\otimes 2}.
\label{redu-1}
\end{eqnarray}%
Thus, via Eq. (\ref{redu-1}), Eq. (\ref{eq-2}) reduces to
\begin{eqnarray}
&&Q_{1}Q_{1}^{t}  \notag \\
&=&[T(\det \mathcal{A}_{q_{1}}\det \mathcal{A}_{q_{2}})\upsilon ^{\otimes
2}T^{t}]\otimes \cdots \\
&&\otimes \lbrack T(\det A_{q_{2n-1}}\det A_{q_{2n}})\upsilon ^{\otimes
2}T^{t}]  \notag \\
&=&[(\det \mathcal{A}_{q_{1}}\det \mathcal{A}_{q_{2}})T\upsilon ^{\otimes
2}T^{t}]\otimes \cdots \\
&&\otimes \lbrack (\det A_{q_{2n-1}}\det A_{q_{2n}})T\upsilon ^{\otimes
2}T^{t}]  \notag \\
&=&(\Pi _{i=1}^{2n}\det A_{q_{i}})(T\upsilon ^{\otimes 2}T^{t}\otimes \cdots
\otimes T\upsilon ^{\otimes 2}T^{t}).  \label{redu-2}
\end{eqnarray}%
One can check that
\begin{equation}
T\upsilon ^{\otimes 2}T^{t}=I_{4}.  \label{redu-3}
\end{equation}%
Thus, from Eqs. (\ref{redu-2}, \ref{redu-3}), we obtain
\begin{equation}
Q_{1}Q_{1}^{t}=(\Pi _{i=1}^{2n}\det A_{q_{i}})I_{2^{2n}}.  \label{orth-1}
\end{equation}%
A calculation also yields
\begin{equation}
Q_{1}Q_{1}^{t}=Q_{1}^{t}Q_{1}.
\end{equation}

Similarly,
\begin{equation}
Q_{2}Q_{2}^{t}=Q_{2}^{t}Q_{2}=(\Pi _{i=1}^{2n}\det A_{q_{2n+i}})I_{2^{2n}}.
\label{orth-2}
\end{equation}

\section*{Appendix B Proportional relations}


\setcounter{equation}{0} \renewcommand{\theequation}{B\arabic{equation}}

Let
\begin{eqnarray}
M_{2n} &=&\left(
\begin{array}{cc}
0 & m \\
m^{t} & 0%
\end{array}%
\right) , \\
D_{2n} &=&\left(
\begin{array}{cc}
0 & hm \\
gm^{t} & 0%
\end{array}%
\right) ,
\end{eqnarray}
where $m$ is an $n$ by $n$ matrix and $g$ and $h$ are non-zero complex
numbers.

Property (1). $\det (\lambda I-D_{2n})=\det (\lambda ^{2}I-ghmm^{t})$. Let $%
\lambda ^{2}=\sigma $. Then, we obtain $\det (\sigma I-ghmm^{t})$. Let $%
a^{2} $ be an eigenvalue of $mm^{t}$. Then, $gha^{2}$ is an eigenvalue of $%
ghmm^{t} $, $\pm a$ are eigenvalues of $M_{2n}$,\ and $\pm \sqrt{gh}a$ are
eigenvalues of $D_{2n}$. Therefore, spectra of $D_{2n}$ and $M_{2n}$ are
proportional.

Property (2.1). Let
\begin{equation}
V=\left(
\begin{array}{c}
v^{\prime } \\
v^{\prime \prime }%
\end{array}%
\right) ,  \label{vector-1}
\end{equation}%
where $v^{\prime }$ and $v^{\prime \prime }$ are $n\times 1$ vectors, be an
eigenvector of $M_{2n}$ corresponding to the eigenvalue $\lambda \neq 0$.
Then,
\begin{equation}
M_{2n}V=\lambda V.  \label{vec-2}
\end{equation}%
Let
\begin{equation}
W=\left(
\begin{array}{c}
\sqrt{h/g}v^{\prime } \\
v^{\prime \prime }%
\end{array}%
\right) .  \label{vec-3}
\end{equation}%
Then, via Eq. (\ref{vec-2}) one can check that
\begin{equation}
D_{2n}W=\sqrt{gh}\lambda W.  \label{vec-4}
\end{equation}%
It means that $W$ is an eigenvector of $D_{2n}$ corresponding to the
eigenvalue $\sqrt{gh}\lambda $.

One can see that if there are $s$ linearly independent eigenvectors $\left(
\begin{array}{c}
v_{i}^{\prime } \\
v_{i}^{\prime \prime }%
\end{array}%
\right) $, $i=1,2,\cdots ,s$, corresponding to the eigenvalue $\lambda $ of $%
M_{2n}$, then there are $s$ linearly independent eigenvectors $\left(
\begin{array}{c}
\sqrt{h/g}v_{i}^{\prime } \\
v_{i}^{\prime \prime }%
\end{array}%
\right) $, $i=1,2,\cdots ,s$, corresponding to the eigenvalue $\sqrt{gh}%
\lambda $ of $D_{2n}$. Therefore, the eigenvalue $\lambda $ of $M_{2n}$\ and
the eigenvalue$\sqrt{gh}\lambda $ of $D_{2n}$\ possess the same geometry
multiplicity. It implies that $M_{2n}$ and $D_{2n}$ have the same number of
JBs corresponding to the eigenvalues $\lambda $ and $\sqrt{gh}\lambda $.

Property (2.2). Let $V$ be an eigenvector of $M_{2n}$ corresponding to the
zero-eigenvalue. Then, one can check that $V$ is also an eigenvector of $%
D_{2n}$\ corresponding to the zero-eigenvalue. It means that the
zero-eigenvalue of $M_{2n}$ and the zero-eigenvalue of $D_{2n}$ possess the
same eigenspace and of course the same geometry multiplicity. Thus, $M_{2n}$
and $D_{2n}$ have the same number of JBs corresponding to the
zero-eigenvalue.

Property (3). $M_{2n}$ has a JB with\ the size of $r$ corresponding to the
eigenvalue $\lambda $ if and only if $D_{2n}$ has a JB with\ the size of $r$
corresponding to the eigenvalue $\sqrt{gh}\lambda $. The property is also
true when $\lambda =0$.

Suppose that $M_{2n}$ has a JB with\ the size of $r$ corresponding to the
eigenvalue $\lambda $ ($\lambda $ may be zero). Then, there exists a Jordan
chain with the size of $r$ corresponding to the eigenvalue $\lambda $ \cite%
{Richard}. Let the Jordan chain be
\begin{equation}
v_{i}=\left(
\begin{array}{c}
v_{i}^{\prime } \\
v_{i}^{\prime \prime }%
\end{array}%
\right) ,i=1,2,\cdots ,r,
\end{equation}%
where $v_{i}^{\prime }$ and $v_{i}^{\prime \prime }$ are $n\times 1$
vectors, $v_{1}$ is the eigenvector, and $v_{i}$ satisfy
\begin{equation}
\left( M_{2n}-\lambda I_{2n}\right) v_{i}=v_{i-1},i=2,\cdots ,r.  \label{M-1}
\end{equation}

From the Jordan chain, we construct the following chain:%
\begin{eqnarray}
z_{1} &=&\left(
\begin{array}{c}
\sqrt{h/g}v_{1}^{\prime } \\
v_{1}^{\prime \prime }%
\end{array}%
\right) , \\
z_{i} &=&\left(
\begin{array}{c}
\frac{1}{g(\sqrt{gh})^{i-2}}v_{i}^{\prime } \\
\frac{1}{(\sqrt{gh})^{i-1}}v_{i}^{\prime \prime }%
\end{array}%
\right) ,i=2,\cdots ,r.
\end{eqnarray}%
One can test that $z_{1}$ is an eigenvector of $D_{2n}$ corresponding to the
eigenvalue $\sqrt{gh}\lambda $. A calculation yields that
\begin{eqnarray}
D_{2n}z_{i} &=&\left(
\begin{array}{cc}
0 & hm \\
gm^{t} & 0%
\end{array}%
\right) \left(
\begin{array}{c}
\frac{1}{g(\sqrt{gh})^{i-2}}v_{i}^{\prime } \\
\frac{1}{(\sqrt{gh})^{i-1}}v_{i}^{\prime \prime }%
\end{array}%
\right)  \notag \\
&=&\left(
\begin{array}{c}
\frac{h}{(\sqrt{gh})^{i-1}}mv_{i}^{\prime \prime } \\
\frac{1}{(\sqrt{gh})^{i-2}}m^{t}v_{i}^{\prime }%
\end{array}%
\right) \\
&=&\left(
\begin{array}{c}
\frac{gh}{g(\sqrt{gh})^{i-1}}mv_{i}^{\prime \prime } \\
\frac{1}{(\sqrt{gh})^{i-2}}m^{t}v_{i}^{\prime }%
\end{array}%
\right) \\
&=&\left(
\begin{array}{c}
\frac{1}{g(\sqrt{gh})^{i-3}}mv_{i}^{\prime \prime } \\
\frac{1}{(\sqrt{gh})^{i-2}}m^{t}v_{i}^{\prime }%
\end{array}%
\right) ,  \label{D-1}
\end{eqnarray}%
\begin{eqnarray}
\sqrt{gh}\lambda I_{2n}z_{i} &=&\sqrt{gh}\lambda \left(
\begin{array}{c}
\frac{1}{g(\sqrt{gh})^{i-2}}v_{i}^{\prime } \\
\frac{1}{(\sqrt{gh})^{i-1}}v_{i}^{\prime \prime }%
\end{array}%
\right)  \notag \\
&=&\left(
\begin{array}{c}
\frac{1}{g(\sqrt{gh})^{i-3}}\lambda v_{i}^{\prime } \\
\frac{1}{(\sqrt{gh})^{i-2}}\lambda v_{i}^{\prime \prime }%
\end{array}%
\right) ,  \label{D-2}
\end{eqnarray}

and
\begin{equation}
z_{i-1}=\left(
\begin{array}{c}
\frac{1}{g(\sqrt{gh})^{i-3}}v_{i-1}^{\prime } \\
\frac{1}{(\sqrt{gh})^{i-2}}v_{i-1}^{\prime \prime }%
\end{array}%
\right) .  \label{D-3}
\end{equation}

From Eqs. (\ref{M-1}, \ref{D-1}, \ref{D-2}, \ref{D-3}), we can show that
\begin{equation}
\left( D_{2n}-\sqrt{gh}\lambda I_{2n}\right) z_{i}=z_{i-1},i=2,\cdots ,r.
\end{equation}%
Thus, we obtain a Jordan chain $z_{1}$, $\cdots $, $z_{r}$ corresponding to
the eigenvalue $\sqrt{gh}\lambda $ of $D_{2n}$. It means that the two Jordan
chains have the same size. Note that the Jordan chain $z_{1}$, $\cdots $, $%
z_{r}$ corresponds to the JB of size $r$ corresponding to the eigenvalue $%
\sqrt{gh}\lambda $ of $D_{2n}$. Conversely, it is also true.

\section*{\protect\bigskip Appendix C Properties of the matrix $M_{2n}$}


\setcounter{equation}{0} \renewcommand{\theequation}{C\arabic{equation}}

Let%
\begin{equation}
M_{2n}=\left(
\begin{array}{cc}
0 & m \\
m^{t} & 0%
\end{array}%
\right) ,
\end{equation}%
\ where $m$ is an $n$ by $n$ matrix. We calculate the characteristic
polynomial of $M_{2n}$ below.

\begin{eqnarray}
&&\det (\lambda I_{2n}-M_{2n})  \notag \\
&=&\det \left( \lambda ^{2}I_{n}-mm^{t}\right) =\det \left( \lambda
^{2}I_{n}-m^{t}m\right) .  \label{e-value}
\end{eqnarray}

Eq. (\ref{e-value}) leads to the following property 1.

\textit{Property 1.}

1.1. $\lambda $ is an eigenvalue of $M_{2n}$ if and only if $\lambda ^{2}$
is an eigenvalue of $m^{t}m$\ and $mm^{t}$, respectively. Thus, the non-zero
eigenvalues of $M_{2n}$ are $\pm \lambda _{i}$, $i=1,2,$ $...$.

1.2. The AM of the zero-eigenvalue\ of $M_{2n}$ is even.

\textit{Property 2.} If $V$ in Eq. (\ref{vector-1})\ is an eigenvector of $%
M_{2n}$ corresponding to the zero-eigenvalue, then $V_{1}=\left(
\begin{array}{c}
v^{\prime } \\
0%
\end{array}%
\right) $ (if $v^{\prime }\neq 0$) and $V_{2}=\left(
\begin{array}{c}
0 \\
v^{\prime \prime }%
\end{array}%
\right) $ (if $v^{\prime \prime }\neq 0$) are also eigenvectors of $M_{2n}$
corresponding to the zero-eigenvalue. Clearly, $V$ is a linear combination
of $V_{1}$ and $V_{2}$, i.e. $V=V_{1}+V_{2}$.

Proof. From that $M_{2n}V=0$, we obtain
\begin{eqnarray}
mv^{\prime \prime } &=&0,  \label{RR-1} \\
m^{t}v^{\prime } &=&0.  \label{RR-2}
\end{eqnarray}%
It is easy to verify that $V_{1}=\left(
\begin{array}{c}
v^{\prime } \\
0%
\end{array}%
\right) $ (if $v^{\prime }\neq 0$) and $V_{2}=\left(
\begin{array}{c}
0 \\
v^{\prime \prime }%
\end{array}%
\right) $ (if $v^{\prime \prime }\neq 0$) are also eigenvectors of $M_{2n}$
corresponding to the zero-eigenvalue.

\textit{Property 3.} The GM of the zero-eigenvalue of $M_{2n}$ is $%
2(n-rk(m)) $, where $rk$ stands for \textquotedblleft rank\textquotedblright
. Thus, there are $2(n-rk(m))$ JBs corresponding to the zero-eigenvalue of $%
M_{2n}$.

Proof. From the linear algebra, it is easy to see that Property 3 holds. We
want to prove it differently next. From \cite{Richard}, we know that the
generalized eigenvector of rank $1$ is just an eigenvector. For $M_{2n}$,
let $\chi _{1}$ be the number of linear independent generalized eigenvectors
of rank $1$ corresponding to the zero-eigenvalue. Then, from \cite{Richard}
\begin{equation}
\chi _{1}=2n-rk(M_{2n}).
\end{equation}%
It is easy to see that $rk(M_{2n})=2\ast rk(m)$.

\textit{Property 4.} A basis of the zero-eigenspace of $M_{2n}$ can be
obtained via the bases of the zero-eigenspaces of $m$ and $m^{t}$ as
follows. Let $v_{1}^{\prime }$, $v_{2}^{\prime }$, $\cdots $, $%
v_{n-rk(m)}^{\prime }$ be all the linearly independent eigenvectors of $%
m^{t} $ corresponding to the zero-eigenvalue and $v_{1}^{\prime \prime }$, $%
v_{2}^{\prime \prime }$, $\cdots $, $v_{n-rk(R)}^{\prime \prime }$ be all
the linearly independent eigenvectors of $m$ corresponding to the
zero-eigenvalue. Then,
\begin{eqnarray}
&&\{\left(
\begin{array}{c}
v_{1}^{\prime } \\
0%
\end{array}%
\right) ,\cdots ,\left(
\begin{array}{c}
v_{n-rk(m)}^{\prime } \\
0%
\end{array}%
\right) ,  \notag \\
&&\left(
\begin{array}{c}
0 \\
v_{1}^{\prime \prime }%
\end{array}%
\right) ,\cdots ,\left(
\begin{array}{c}
0 \\
v_{n-rk(m)}^{\prime \prime }%
\end{array}%
\right) \}
\end{eqnarray}%
is a basis of the zero-eigenspace of $M_{2n}$.

Proof. Let $V$ in Eq. (\ref{vector-1}) be an eigenvector of $M_{2n}$
corresponding to the zero-eigenvalue. Then, by Eqs. (\ref{RR-1}, \ref{RR-2}%
), $v^{\prime \prime }$ is an eigenvector of $m$ corresponding to the
zero-eigenvalue if $v^{\prime \prime }\neq 0$ and $v^{\prime }$ is an
eigenvector of $m^{t}$ corresponding to the zero-eigenvalue if $v^{\prime
}\neq 0$. Conversely, if $v^{\prime }$ (resp. $v^{\prime \prime }$) is an
eigenvector of $m^{t}$ (resp. $m$) corresponding to the zero-eigenvalue,
then $\left(
\begin{array}{c}
v^{\prime } \\
0%
\end{array}%
\right) $ (resp. $\left(
\begin{array}{c}
0 \\
v^{\prime \prime }%
\end{array}%
\right) $) is an eigenvector of $M_{2n}$ corresponding to the
zero-eigenvalue. From Eqs. (\ref{RR-1}, \ref{RR-2}), we know that $m$ and $%
m^{t}$\ have $n-rk(m)$ linearly independent eigenvectors corresponding to
the zero-eigenvalue, respectively. Thus, Property 4 holds and we have
Property 3 again.

\textit{Property 5.1}. For $M_{2n}$, let $\chi _{\ell }$ be the number of
linear independent generalized eigenvectors of rank $\ell $ corresponding to
the zero-eigenvalue \cite{Richard}. Then, $\chi _{2k}+\chi _{2k+1}$, where $%
k\geq 1$, must be even.

Proof. From \cite{Richard},
\begin{equation}
\chi _{2k}=rk(M_{2n}^{2k-1})-rk(M_{2n}^{2k})
\end{equation}%
and
\begin{equation}
\chi _{2k+1}=rk(M_{2n}^{2k})-rk(M_{2n}^{2k+1}),
\end{equation}%
where $k\geq 1$. Then,
\begin{equation}
\chi _{2k}+\chi _{2k+1}=rk(M_{2n}^{2k-1})-rk(M_{2n}^{2k+1}).
\end{equation}%
Let us compute $M_{2n}^{2k+1}$ next.
\begin{eqnarray}
M_{2n}^{3} &=&\left(
\begin{array}{cc}
& mm^{t}m \\
m^{t}mm^{t} &
\end{array}%
\right) , \\
M_{2n}^{2k+1} &=&\left(
\begin{array}{cc}
& M^{\prime } \\
(M^{\prime })^{t} &
\end{array}%
\right) ,
\end{eqnarray}%
where $M^{\prime }=mm^{t}mm^{t}\cdots mm^{t}m$.

It is easy to check that $rk(M_{2n}^{2k+1})=2\ast rk(M^{\prime })$.
Similarly, the number $rk(M_{2n}^{2k-1})$ is even. Therefore, $\chi
_{2k}+\chi _{2k+1}$ is even. Specially, $\chi _{2}+\chi _{3}$ is even.

\textit{Property 5.2.} The number of the occurrences of JBs with the same
odd size\ corresponding to the zero-eigenvalue of $M_{2n}$ may be even or
odd.

Proof. For the JB $J_{2k+1}(0)$ corresponding to the eigenvector $x_{1}$,
there is a Jordan chain $x_{1}$, $x_{2}$, $\cdots $, $x_{2k+1}$
corresponding to the zero-eigenvalue, where $x_{i}$ is the generalized
eigenvector of rank $i$ of $M_{2n}$. Clearly, $x_{2j}$\ is the generalized
eigenvector of rank $2j$ and $x_{2j+1}$ is the one of rank $2j+1$, where $%
j=1,\cdots ,k$. Thus, the chain adds 1 to $\chi _{2j}$ and 1 to $\chi
_{2j+1} $, respectively, $j=1,\cdots ,k$. That is, the chain adds 2 to the
number $\chi _{2j}+\chi _{2j+1}$, $j=1,\cdots ,k$. One can know that any
number of occurrences of JBs with the same odd size\ corresponding to the
zero-eigenvalue will not change the parity of $\chi _{2k}+\chi _{2k+1}$.
Therefore, Property 5.2 holds.

\textit{Property 5.3.} The number of the occurrences of the JBs with the
same even size corresponding to the zero-eigenvalue must be even.

Proof. For the JB $J_{2k}(0)$ with $k\geq 1$ corresponding to the
eigenvector $y_{1}$, there is a Jordan chain $y_{1}$, $y_{2}$, $\cdots $, $%
y_{2k}$ corresponding to the zero-eigenvalue, where $y_{i}$ is the
generalized eigenvector of rank $i$ of $M_{2n}$. Thus, $y_{2j}$ is the
generalized eigenvector of rank $2j$ while $y_{2j+1}$ is the one of rank $%
2j+1$, where $j=1,\cdots ,k-1$. Thus, the chain adds 1 to $\chi _{2j}$ and 1
to $\chi _{2j+1}$ respectively, $j=1,\cdots ,k-1$.

Clearly, $y_{2k}$ is the generalized eigenvector of rank $2k$. Thus, it adds
1 to $\chi _{2k}$. But the chain does not include the generalized
eigenvector of rank $2k+1$. Thus, it adds 0 to $\chi _{2k+1}$. It means that
the chain will change the parity of $\chi _{2k}+\chi _{2k+1}$.

Accordingly, for the $2l+1$ occurrences of the JB $J_{2k}(0)$ with $k\geq 1$%
, the corresponding $2l+1$ Jordan chains include $2l+1$ generalized
eigenvectors of the same rank $2k$ but the chains do not have any
generalized eigenvector of rank $2k+1$. Thus, in light of Property 5.2, the
number $\chi _{2k}+\chi _{2k+1}$ will be an odd number. It does not satisfy
Property 5.1.

For the $2l$\ occurrences of the JB $J_{2k}(0)$ with $k\geq 1$, the
corresponding $2l$ Jordan chains include $2l$ generalized eigenvectors of
the same rank $2k$ but the chains do not have any generalized eigenvector of
rank $2k+1$. Thus, in light of Property 5.2, the number $\chi _{2k}+\chi
_{2k+1}$ will be an even number.

One can see that $\chi _{2k}+\chi _{2k+1}$ is even permits that the size of
a JB with the zero-eigenvalue is odd or even.

For example, a calculation shows that for four qubits, $\Phi _{8}$ has the
SJNFs $J_{4}(0)^{\oplus 2}$, $J_{2}(0)^{\oplus 2}\oplus J_{1}(0)^{\oplus 4}$%
, $J_{2}(0)^{\oplus 2}\oplus J_{3}(0)\oplus J_{1}(0)$, $J_{2}(0)^{\oplus 4}$
for the states $L_{a_{4}}(a=0)$, $L_{abc_{2}}(a=b=c=0)$, $L_{a_{2}0_{3\oplus
1}}(a=0)$, $L_{a_{2}b_{2}}(a=b=0)$, respectively. In detail, $J_{4}(0)$
occurs twice, $J_{2}(0)$ occurs twice, twice, and for four times in the
above SJNFs. See Table II. For these SJNFs, $\chi _{2k}+\chi _{2k+1}$ is
even.

One can know that $\Phi _{8}$ does not have SJNFs $\pm \lambda
J_{2}(0)J_{4}(0)$, $J_{2}(0)J_{6}(0)$ or $00J_{2}(0)J_{4}(0)$ because for
these SJNFs $\chi _{2}+\chi _{3}$ is odd. Note that $J_{2}(0)$, $J_{4}(0)$,
and $J_{6}(0)$ occur once in the above different SJNFs.

\textit{Property 6.}

Let $V$ in Eq. (\ref{vector-1})\ be an eigenvector of $M_{2n}$ corresponding
to the non-zero eigenvalue $\lambda $. Then, $v^{\prime }\neq 0$ and $%
v^{\prime \prime }\neq 0$.

Proof.\ From the equation $\left( M_{2n}-\lambda I_{2n}\right) V=0$, we
obtain
\begin{eqnarray}
mv^{\prime \prime } &=&\lambda v^{\prime }\text{, }  \label{eigen-1} \\
m^{t}v^{\prime } &=&\lambda v^{\prime \prime }\text{.}  \label{eigen-2}
\end{eqnarray}%
Then from Eqs. (\ref{eigen-1}, \ref{eigen-2}), it is easy to show that $%
v^{\prime }\neq 0$ and $v^{\prime \prime }\neq 0$. In other words, the
vectors of the forms $\left(
\begin{array}{c}
v^{\prime } \\
0%
\end{array}%
\right) $ or $\left(
\begin{array}{c}
0 \\
v^{\prime \prime }%
\end{array}%
\right) $ are not eigenvectors of $M_{2n}$ corresponding to non-zero
eigenvalues.

\textit{Property 7.} The GMs of the non-zero eigenvalues $\pm \lambda $ of $%
M_{2n}$ both are $n-rk(m^{t}m-\lambda ^{2}I_{n})$. Thus, there are $%
n-rk(m^{t}m-\lambda ^{2}I_{n})$ JBs corresponding to the non-zero
eigenvalues $\pm \lambda $ of $M_{2n}$, respectively.

Proof. Let $\chi _{1}(\lambda )$\ (resp. $\chi _{1}(-\lambda )$) be the
number of linear independent generalized eigenvectors of rank $1$
corresponding to the non-zero eigenvalue $\lambda $ (resp. $-\lambda $). One
can know that $\chi _{1}(\lambda )$\ (resp. $\chi _{1}(-\lambda )$) is just
the GMs of the non-zero eigenvalues $\lambda $ (resp. $-\lambda $) of $%
M_{2n} $. Then, from \cite{Richard}

\begin{eqnarray}
&&\chi _{1}(\lambda )=2n-rk(M_{2n}-\lambda I_{2n})  \notag \\
&=&2n-rk\left(
\begin{array}{cc}
-\lambda I_{n} & m \\
m^{t} & -\lambda I_{n}%
\end{array}%
\right) .  \label{eq-3}
\end{eqnarray}

\begin{eqnarray}
\chi _{1}(-\lambda ) &=&2n-rk(M_{2n}+\lambda I_{2n})  \notag \\
&=&2n-rk\left(
\begin{array}{cc}
\lambda I_{n} & m \\
m^{t} & \lambda I_{n}%
\end{array}%
\right) .  \label{eq-4}
\end{eqnarray}

To calculate the ranks of the matrices $\left(
\begin{array}{cc}
-\lambda I_{n} & m \\
m^{t} & -\lambda I_{n}%
\end{array}%
\right) $ and $\left(
\begin{array}{cc}
\lambda I_{n} & m \\
m^{t} & \lambda I_{n}%
\end{array}%
\right) $, we do the following operations:
\begin{eqnarray}
&&\left(
\begin{array}{cc}
I_{n} & 0 \\
\frac{1}{\lambda }m^{t} & I_{n}%
\end{array}%
\right) \left(
\begin{array}{cc}
-\lambda I_{n} & m \\
m^{t} & -\lambda I_{n}%
\end{array}%
\right)  \notag \\
&=&\left(
\begin{array}{cc}
-\lambda I_{n} & m \\
0 & \frac{1}{\lambda }m^{t}m-\lambda I_{n}%
\end{array}%
\right)  \label{eq-5}
\end{eqnarray}

and
\begin{eqnarray}
&&\left(
\begin{array}{cc}
-I_{n} & 0 \\
\frac{1}{\lambda }m^{t} & -I_{n}%
\end{array}%
\right) \left(
\begin{array}{cc}
\lambda I_{n} & m \\
m^{t} & \lambda I_{n}%
\end{array}%
\right)  \notag \\
&=&\left(
\begin{array}{cc}
-\lambda I_{n} & -m \\
0 & \frac{1}{\lambda }m^{t}m-\lambda I_{n}%
\end{array}%
\right) .  \label{eq-6}
\end{eqnarray}

From the linear algebra, since $\left(
\begin{array}{cc}
I_{n} & 0 \\
\frac{1}{\lambda }m^{t} & I_{n}%
\end{array}%
\right) $ is full rank, via Eq. (\ref{eq-5}) we obtain
\begin{eqnarray}
rk\left(
\begin{array}{cc}
-\lambda I_{n} & m \\
m^{t} & -\lambda I_{n}%
\end{array}%
\right) &=&rk\text{ }\left(
\begin{array}{cc}
-\lambda I_{n} & m \\
0 & \frac{1}{\lambda }m^{t}m-\lambda I_{n}%
\end{array}%
\right)  \notag \\
&=&n+rk(\frac{1}{\lambda }m^{t}m-\lambda I_{n})  \notag \\
&=&n+rk(\frac{1}{\lambda }(m^{t}m-\lambda ^{2}I_{n}))  \notag \\
&=&n+rk(m^{t}m-\lambda ^{2}I_{n})  \label{eq-7}
\end{eqnarray}

From the linear algebra, since $\left(
\begin{array}{cc}
-I_{n} & 0 \\
\frac{1}{\lambda }m^{t} & -I_{n}%
\end{array}%
\right) $ is full rank, via Eq. (\ref{eq-6}) we obtain
\begin{eqnarray}
rk\left(
\begin{array}{cc}
\lambda I_{n} & m \\
m^{t} & \lambda I_{n}%
\end{array}%
\right) &=&rk\left(
\begin{array}{cc}
-\lambda I_{n} & -m \\
0 & \frac{1}{\lambda }m^{t}m-\lambda I_{n}%
\end{array}%
\right)  \notag \\
&=&n+rk(\frac{1}{\lambda }m^{t}m-\lambda I_{n})  \notag \\
&=&n+rk(m^{t}m-\lambda ^{2}I_{n}).  \label{eq-8}
\end{eqnarray}

From Eqs. (\ref{eq-3}, \ref{eq-7}), $\chi _{1}(\lambda
)=2n-[n+rk(m^{t}m-\lambda ^{2}I_{n})]=n-rk(m^{t}m-\lambda ^{2}I_{n})$.
Clearly, $m^{t}m-\lambda ^{2}I_{n}$ is a characteristic matrix of $m^{t}m$
in $\lambda ^{2}$. From Eqs. (\ref{eq-4}, \ref{eq-8}), $\chi _{1}(-\lambda
)=n-rk(m^{t}m-\lambda ^{2}I_{n})$. Therefore, $\chi _{1}(\lambda )=\chi
_{1}(-\lambda )$ and then Property 7 holds.

By Property 1.1, when $\lambda $ is an eigenvalue of $M_{2n}$, then $\lambda
^{2}$ is an eigenvalue of $m^{t}m$. It is well known that roots of the
equation $\det (m^{t}m-\lambda ^{2}I_{n})=0$ are eigenvalues of $m^{t}m$.\
Thus, $0\leq rk(m^{t}m-\lambda ^{2}I_{n})<n$ and then $0<\chi _{1}(\lambda
)\leq n$. When $\lambda ^{2}$ is not an eigenvalue of $m^{t}m$, i.e. $%
\lambda $ is not an eigenvalue of $M_{2n}$, then $\det (m^{t}m-\lambda
^{2}I_{n})\neq 0$, i.e. $rk(m^{t}m-\lambda ^{2}I_{n})=n$. Thus, $\chi
_{1}(\lambda )=0$.

\textit{Property 8}. The Jordan chain with the non-zero eigenvalue $\lambda $
corresponding to the eigenvector $\left(
\begin{array}{c}
v_{1}^{\prime } \\
v_{1}^{\prime \prime }%
\end{array}%
\right) $ and the Jordan chain with the non-zero eigenvalue $-\lambda $
corresponding to the eigenvector $\left(
\begin{array}{c}
-v_{1}^{\prime } \\
v_{1}^{\prime \prime }%
\end{array}%
\right) $ have the same size. Thus, their corresponding JBs have the same
size.

Proof. Let
\begin{equation}
v_{i}=\left(
\begin{array}{c}
v_{i}^{\prime } \\
v_{i}^{\prime \prime }%
\end{array}%
\right) ,i=1,2,\cdots ,r,
\end{equation}%
where $v_{i}^{\prime }$ and $v_{i}^{\prime \prime }$ are $n\times 1$
vectors, be a Jordan chain with the non-zero eigenvalue $\lambda $
corresponding to the eigenvector $v_{1}=\left(
\begin{array}{c}
v_{1}^{\prime } \\
v_{1}^{\prime \prime }%
\end{array}%
\right) $. By Property 6, $v_{1}^{\prime }\neq 0$ and $v_{1}^{\prime \prime
}\neq 0$.\ Then, by the definition of Jordan chain \cite{Richard},

\begin{equation}
(M_{2n}-\lambda I_{2n})v_{1}=0
\end{equation}%
and
\begin{equation}
(M_{2n}-\lambda I_{2n})v_{k}=v_{k-1},k\geq 2.
\end{equation}

Let
\begin{eqnarray}
\omega _{1} &=&\left(
\begin{array}{c}
-v_{1}^{\prime } \\
v_{1}^{\prime \prime }%
\end{array}%
\right) , \\
\omega _{2} &=&\left(
\begin{array}{c}
v_{2}^{\prime } \\
-v_{2}^{\prime \prime }%
\end{array}%
\right) , \\
&&\vdots  \notag \\
\omega _{l} &=&(-1)^{l+1}\left(
\begin{array}{c}
-v_{l}^{\prime } \\
v_{l}^{\prime \prime }%
\end{array}%
\right) ,l\geq 2.
\end{eqnarray}%
It is easy to check that
\begin{equation}
(M_{2n}+\lambda I_{2n})\omega _{1}=0
\end{equation}%
and
\begin{equation}
(M_{2n}+\lambda I_{2n})\omega _{k}=\omega _{k-1},k\geq 2.
\end{equation}%
Here, $\omega _{1}$ is an eigenvector of $M_{2n}$ corresponding to $-\lambda
$. Let $s$ be the size\ of the Jordan chain with the non-zero eigenvalue $%
-\lambda $ corresponding to the eigenvector $\omega _{1}$. Clearly, $s\geq r$%
. Conversely, similarly, we can show that $r\geq s$. Thus, $s=r$.

\section*{Appendix D The number of different lists of partitions of AMs}


\setcounter{equation}{0} \renewcommand{\theequation}{D\arabic{equation}}

We define a product of sets $\overline{L}$ and $\overline{M}$ as $\overline{L%
}\times \overline{M}=[\{l,m\}|l\in $ $\overline{l}$ and $m\in \overline{M}]$
and we define that $\{l,m\}$ is an unordered list of partitions. Thus, $%
\{l,m\}=\{m,l\}$. By the definition, $\overline{2}\times \overline{2}%
=[\{(2),(2)\}$, $\{(2),(1,1)\}$, $\{(1,1),(1,1)\}]$. Note that $%
\{(1,1),(2)\}=\{(2),(1,1)\}$.

From Eq. (\ref{SJNF-1}), let
\begin{equation}
\Gamma =\widetilde{2k}\times \overline{\ell _{1}}\times \overline{\ell _{2}}%
\cdots \times \overline{\ell _{s}}.  \label{prod-1}
\end{equation}%
From the above discussion, we consider that $\overline{\ell _{1}}\times
\overline{\ell _{2}}\cdots \times \overline{\ell _{s}}$ is an unordered list
of partitions. Note that some $\ell _{i}$ in a set of AMs $\{2k;\ell
_{1},\ell _{2},\cdots ,\ell _{s}\}$ from Eq. (\ref{Jordan}) may occur twice
or more. For example, $\Phi _{8}$ has the spectrum $\{(\pm \lambda
_{1}{})^{\circledcirc 2},(\pm \lambda _{2}{})^{\circledcirc 2}\}$ and the
set of the AM is $\{0;2,2\}$.

First, let us compute how many different lists of partitions there are from
the product set $\underbrace{\overline{l}\times \cdots \times \overline{l}}%
_{j}$. We consider distributing $j$ indistinguishable balls into $P(l)$
distinguishable boxes. Let $\rho (l,j)=\left(
\begin{array}{c}
j+P(l)-1 \\
j%
\end{array}%
\right) $.\ Thus, there are $\rho (l,j)$ distributing ways without exclusion
\cite{DeGroot}. Via the probability model, $\underbrace{\overline{l}\times
\cdots \times \overline{l}}_{j}$ has $\rho (l,j)$ different lists of
partitions. Specially, $\overline{2}\times \overline{2}$ has $\rho (2,2)$ ($%
=3$) different lists of partitions.

It is easy to check that $\overline{2}\times \overline{3}$ has $P(2)P(3)=6$
different lists of partitions. When $l$, $k$, and $m$ are distinct from each
other, $\overline{l}\times \overline{k}\times \overline{m}$ has $%
P(l)P(k)P(m) $ different lists of partitions.

Let us compute how many different lists of partitions there are from the
product set $\Gamma $\ in Eq. (\ref{prod-1}) for all $k$. For the sake of
clarity, we rewrite $\Gamma $ in Eq. (\ref{prod-1}) as follows:

\begin{equation}
\Gamma =\widetilde{2k}\times \overline{\varkappa _{1}}\times \cdots \times
\overline{\varkappa _{i}}\times \underbrace{\overline{\theta }\times \cdots
\times \overline{\theta }}_{j}\times \cdots \times \underbrace{\overline{%
\varsigma }\times \cdots \times \overline{\varsigma }}_{m},  \label{prod-2}
\end{equation}%
where $\varkappa _{1}$, $\cdots $, $\varkappa _{i}$, $\theta $, $\cdots $,
and $\varsigma $ are different from each other. From Eq. (\ref{prod-2}), for
all $k$\ we obtain
\begin{equation}
\eta =\sum_{k=0}^{2^{2n}}P^{\ast }(2k)\sum_{\varpi }P(\varkappa _{1})\cdots
P(\varkappa _{i})\rho (\theta ,j)\cdots \rho (\varsigma ,m)  \label{set-p-2}
\end{equation}%
different lists of partitions, where $\varpi =\{\varkappa _{1},\cdots
,\varkappa _{i},\underbrace{\theta ,\cdots ,\theta }_{j},\cdots ,\underbrace{%
\varsigma ,\cdots ,\varsigma }_{m}\}$, which is a partition of $2^{2n}-k$
and the second sum is evaluated over all the partitions of $2^{2n}-k$.

To compute $P^{\ast }(2^{2n+1})$,\ from $\widetilde{2^{2n+1}}$ we should
remove the partition $(\underbrace{1,\cdots ,1}_{2^{2n+1}})$, which means
that the SJNF of the $\Phi _{2^{2n+1}}(|\psi \rangle )$\ in Eq. (\ref{coef-1}%
) is the zero matrix and then $\Phi _{2^{2n+1}}(|\psi \rangle )$\ in Eq. (%
\ref{coef-1})\ is the zero matrix. Therefore, in total we obtain $\eta -1$
different lists of partitions of AMs in $(2k;\ell _{1},\ell _{2},\cdots
,\ell _{s})$.

\end{document}